\documentclass[preprint,floats,floatfix,superscriptaddress]{revtex4}

\usepackage{amsbsy}
\usepackage{amssymb}
\usepackage{amsmath}
\input{epsf}
\usepackage{graphicx}
\usepackage{color}

\usepackage[caption=false]{subfig}
\usepackage{float}
\usepackage{subfig}
\usepackage{gensymb}
\usepackage{units}
\usepackage[flushleft]{threeparttable}

\usepackage{aas_macros}

\usepackage{changes}

\usepackage{tikz,xcolor,hyperref}

\definecolor{lime}{HTML}{A6CE39}
\DeclareRobustCommand{\orcidicon}{%
	\begin{tikzpicture}
	\draw[lime, fill=lime] (0,0) 
	circle [radius=0.16] 
	node[white] {{\fontfamily{qag}\selectfont \tiny ID}};
	\draw[white, fill=white] (-0.0625,0.095) 
	circle [radius=0.007];
	\end{tikzpicture}
	\hspace{-2mm}
}

\foreach \x in {A, ..., Z}{%
	\expandafter\xdef\csname orcid\x\endcsname{\noexpand\href{https://orcid.org/\csname orcidauthor\x\endcsname}{\noexpand\orcidicon}}
}

\begin{document}

\title{Instabilities in neutron-star postmerger remnants}

\author{Xiaoyi Xie \orcidA{}}
\email{x.xie@soton.ac.uk}
\affiliation{Mathematical Sciences and STAG Research Centre, University of Southampton, Southampton SO17 1BJ, United Kingdom}
\author{Ian Hawke \orcidB{}}
\affiliation{Mathematical Sciences and STAG Research Centre, University of Southampton, Southampton SO17 1BJ, United Kingdom}
\author{Andrea Passamonti \orcidC{}}
\affiliation{Via Greve 10, 00146, Roma, Italy}
\author{Nils Andersson \orcidD{}}
\affiliation{Mathematical Sciences and STAG Research Centre, University of Southampton, Southampton SO17 1BJ, United Kingdom}

\begin{abstract}
Using nonlinear, fully relativistic, simulations we investigate the dynamics and gravitational-wave signature associated with instabilities in neutron star postmerger remnants. For simplified models of the remnant we establish the presence of an instability in stars with moderate $T/|W|$, the ratio between the kinetic and the gravitational potential energies. 
Detailed analysis of the density oscillation pattern reveals a \emph{local} instability in the inner region of the more realistic differential rotation profile.
We apply Rayleigh's inflection theorem and Fj{\o}rtoft's theorem to analyze the stability criteria concluding that this \emph{inner local} instability originates from a shear instability close to the peak of the angular velocity profile, and that it later evolves into a fast-rotating $m=2$ oscillation pattern.
We discuss the importance of the presence of a  corotation point in the fluid, its connection with the shear instability, and comparisons to the Rossby wave and Papaloizou Pringle instabilities considered in the wider literature.
\end{abstract}

\maketitle

\section{Introduction} \label{sec:intro}

The merger of two neutron stars, following the gravitational-wave driven inspiral of a compact binary system, leads to the formation of a hot, differentially rotating remnant \citep{2017RPPh...80i6901B}. Observations of gravitational waves from such mergers are expected to shed light on the nature of matter under extreme pressures and densities by constraining  the maximum mass allowed by the (hot) matter equation of state (e.g.~see the recent review \citep{2020arXiv200406419B}). This puts the nonlinear dynamics of the merger into focus. It has been established that the gravitational-wave signal has robust features, most likely associated with the fundamental f-mode oscillations of the remnant \cite{2012PhRvL.108a1101B,2014PhRvL.113i1104T,2015PhRvL.115i1101B,2016EPJA...52...56B,2016PhRvD..93l4051R}. Less well understood--partly because long-term postmerger simulations are prohibitively expensive (see e.g.~\citep{2019PhRvD.100b3005C})--are  issues relating to the long-term survival of the remnant. The final fate of the hot remnant depends more or less directly on the involved masses (assuming that only a small amount of matter is ejected during the merger), but the time cale on which the object settles down, or collapses to form a black hole, depends on complex issues involving both dissipative mechanisms (in particular associated with neutrino emission) and magnetic field dynamics (see \cite{2019ApJ...880L..15M} for a recent qualitative discussion). In essence, a better understanding of merger remnant dynamics requires progress on both computational issues--how do we track the medium to long-term evolution beyond the merger?--and the physics implementation--what are the most important aspects and how do we implement them in nonlinear simulations? 

This study makes a modest contribution to the discussion by focusing on the impact of the differential rotation of the remnant. This is relevant for several reasons. First, it has been established that the rotation profile of a merger remnant is quite different from that commonly assumed in studies of differentially rotating stars \citep{1986ApJS...61..479H, Komatsu1989MNRASA, Komatsu1989MNRASB}.  Most previous work on differentially rotating stars assumes that the differential rotation profile corresponds to constant specific angular momentum, leading to a profile for the angular velocity, $\Omega$, that falls off  away from the rotation axis. In contrast, merger simulations suggest that the  profile should be fairly flat close to the rotation axis, rising towards a maximum at some point in the remnant beyond which it tapers off towards the profile expected for a Keplerian disk \citep{2015PhRvD..91f4027K, 2017PhRvD..96d3004H, 2017PhRvD..96d3019K, 2020PhRvD.101f4052D}. It seems relevant to ask how this rotation profile impacts on the dynamics, e.g.\ the expected f-mode oscillations. The second relevant aspect concerns the stability of these oscillations. It is well established that the oscillations of differentially rotating stars may be (dynamically) unstable already at rather modest levels of rotation (typically expressed in terms of the ratio of kinetic energy to (the magnitude of the) gravitational potential energy, $T/|W|$). Previous work shows that differentially rotating neutron stars  become dynamically unstable when $T/|W|\ge 0.24$ \citep{2000ApJ...542..453S,2007PhRvD..75d4023B}, but instabilities have  been observed for values as small as $T/|W|\approx 0.01$ for somewhat extreme rotation profiles \cite{2003MNRAS.343..619S}.  The question then is, should we expect such low-$T/|W|$ instabilities to be active in merger remnants? It seems a possibility worth considering, given that we do not yet have a clear understanding of the impact of the actual differential rotation profile for mergers. An initial exploration of the issue, in the context of Newtonian gravity and linear perturbations, gives an affirmative answer to the question \citep{2020arXiv200310198P}. Meanwhile, the work we present here considers the problem in full nonlinear general relativity. This is important because the differential rotation profile evolves as the system settles down, a feature that cannot be represented perturbatively. The question is if (and if so, how) this impacts on the development of an instability. Finally, a question worth asking is whether the difference in the rotation profile introduces new features not found before.

We explore these issues by carrying out numerical simulations of rapidly rotating neutron stars described by two different rotation profiles at (relatively) modest values of  $T/|W|$. We do not consider more extreme cases as they would mainly be of academic interest. 
In Sec. \ref{sec:method}, we summarize the initial data and numerical setup for the simulations. We also lay out the tools utilized in the data analysis. The results are reported in Sec. \ref{sec:results}.

\section{Methodology}\label{sec:method}

\subsection{Numerical setup}

We use the open source code \emph{RNSID} \cite{1995ApJ...444..306S} to construct the rotating initial data, representing stationary equilibrium solutions of axisymmetric  relativistic neutron stars without magnetic fields. We assume that the line element for an axisymmetric and stationary relativistic space-time has the form
\begin{equation}\label{eq:metric}
  ds^2 = -e^{2\nu} dt^2 + e^{2\alpha}(dr^2+r^2d\theta^2)+e^{2\beta} r^2\sin^2\theta( d\phi - \omega dt)^2\ ,
\end{equation}
where $\nu, \alpha,\beta$, and $\omega$ are space dependent metric functions. In generating the equilibrium models,
in the barotropic case, the integrability condition requires that the specific angular momentum measured by the proper time of matter is a function of $\Omega$ \cite{1970ApJ...162...71B,1976ApJ...204..200B,1989MNRAS.237..355K},
\begin{equation}\label{eq:j_rel}
 u^tu_\phi = \frac{v^2}{(1-v^2)(\Omega-\omega)} =  j(\Omega),
\end{equation}
where $\Omega$ is the angular velocity of the matter measured from infinity, and $v=(\Omega-\omega)r\rm{sin}\theta e^{\beta-\nu}$
is the proper velocity with respect to a zero angular momentum observer. The rotation law $j(\Omega)$ used in most of previous work is the so-called $j$-constant law,
\begin{eqnarray}\label{eq:jconst}
  j(\Omega) &=& A^2(\Omega_c -\Omega)\ , \\
  \Omega &=& \Omega_c\left(1- \frac{j}{\Omega_c A^2}\right),
\end{eqnarray}
where $A$ is a positive constant and $\Omega_c$ is the angular velocity at the center. In the Newtonian limit, $j = \Omega \varpi^2$, where $\varpi=r\sin\theta$ is the radial distance from the rotation axis. The rotation law from Eq. \eqref{eq:jconst} can be rewritten as
\begin{equation}
  \Omega/\Omega_c = A^2/(A^2+\varpi^2).
\end{equation}
When $A\to \infty$, it approaches a rigid rotation, while it becomes a $j$-constant rotation when $A\to 0$. That is, in this limit the specific angular momentum is constant in space (see e.g.\ \citet{1985A&A...146..260E}). In relativity, the specific angular momentum $j(\Omega)$ is related to the metric potentials through Eq. \eqref{eq:j_rel}, so the ``$j$-constant'' law becomes:
\begin{equation}
  \Omega_c - \Omega = \frac{1}{\hat{A}^2R_e^2}\left[\frac{(\Omega-\omega)r^2\rm{sin}^2\theta e^{2(\beta-\nu)}}{1-(\Omega-\omega)^2r^2\rm{sin}^2\theta e^{2(\beta-\nu)}}\right]
\end{equation}
where $R_e$ is the coordinate equatorial stellar radius and the coefficient $\hat{A} = A/R_e$ is a measure of the degree of differential rotation.

In order to investigate the dynamics of hypermassive neutron stars  formed after binary neutron star  mergers, the \emph{RNSID} code has been modified to generate representative initial data based on a different differential rotation law introduced by \citet{2017PhRvD..96j3011U}:
\begin{equation}\label{eq:uryu}
  \Omega = \Omega_c\left[1+\left(\frac{j}{\Omega_c B^2}\right)^p\right]\left(1-\frac{j}{\Omega_c A^2}\right),
\end{equation}
where $p$, $A$ and $B$ are parameters that adjust the rotation profile 
and $j$ is given in Eq. \eqref{eq:j_rel} (for examples of different rotation profiles obtained from this prescription, see Fig. 1 in \citet{2020arXiv200310198P}).

The initial data are then evolved using the public domain  {\em Einstein Toolkit} code \citep{2012CQGra..29k5001L}. To evolve the fluid, we use the \emph{GRHydro} module \citep{2014CQGra..31a5005M}, together with the piecewise parabolic reconstruction method \citep{1984JCoPh..54..174CU} and the Marquina flux formula \citep{1999ApJS..122..151A}. We evolve Einstein's equations in the CCZ4 formulation \cite{2012PhRvD..85f4040A}. A fourth-order, conservative Runge-Kutta scheme is used for the time evolution. Both the Einstein and the hydrodynamics equations are solved on a Cartesian grid using the adaptive mesh-refinement approach provided by the \emph{Carpet} driver \citep{2004CQGra..21.1465S}. Seven levels of refinement are used to cover the simulation domain. We adopt units such that $c=M_{\odot}=G=1$  for the simulations. In these code units, the boundary of each refinement level is located at 307.2, 64, 26, 13.6, 8.0, 4.0, and 2.4, respectively. The outermost boundary of the domain is set at $\sim 307\,M_{\odot}(\approx 460\ \rm{km})$, with a resolution of $3.2\,M_{\odot}$ and the finest refinement level has a resolution $dx=0.05\,M_{\odot}(\approx 74\ \rm{m})$. A  $z$-symmetry is imposed for the numerical grid. The density of the surrounding medium (the atmosphere) has been set, relative to the initial central density $\rho_{0, c}$, to a low value $\rho_{\rm{atm}}/\rho_{0,c}\sim 10^{-9}$, leading to $\rho_{\rm{atm}}\sim 6\times 10^5 \rm{g/cm}^3$. 

Finally, we employ refluxing techniques to correct the numerical fluxes across different levels of mesh refinement \citep{2010ApJS..186..308C}. The combination of the refluxing algorithm with a low atmosphere density and the CCZ4 formulation reduces position drift of the rotating profile.  This is important, as it is known that differentially rotating stars may develop spiral instabilities \citep{2006ApJ...651.1068O,2015PhRvD..92l1502P,2016PhRvD..94f4011R,2016CQGra..33x4004E}. As the presence of such modes may be obscured by the numerical code not perfectly conserving linear momentum, it is important to suppress any position drift during the simulation.

 \subsection{Initial data}
 
As our main interest  is in qualitative differences and how these manifest themselves in the evolution of the system,  we carry out simulations for two models. The first uses the  standard $j$-constant rotation law with the dimensionless parameter $\hat{A} $ set to 1. The second model represents the rotation law (\ref{eq:uryu}) with $p=1, \hat A=1$ and $\hat B=0.5$, where $\hat B=B/R_e$.  
In the following, we refer to this as the Ury{\={u}} model to distinguish it from the $j$-constant case. Numerical simulations of merging binary neutron stars show that the rotational profile of the late stage hypermassive neutron star (HMNS) contains a slowly rotating core and an extended envelope rotating close to the Keplerian velocity \citep{PhysRevD.91.064027,2016PhRvD..94d4060K,2017PhRvD..96d3004H}. The Ury{\={u}} model adopted in this study describes the main features of the azimuthally averaged angular-velocity profile of the HMNS found in merger simulations. The focus on two specific models may seem overly restrictive, but the detailed analysis we provide would not be possible for the wider parameter space. Once we have established the tools one may consider  a more exhaustive parameter survey.  We leave this for future work. 

\begin{figure}[!ht]
  \centering
    \includegraphics[clip,width=0.7\columnwidth]
                    {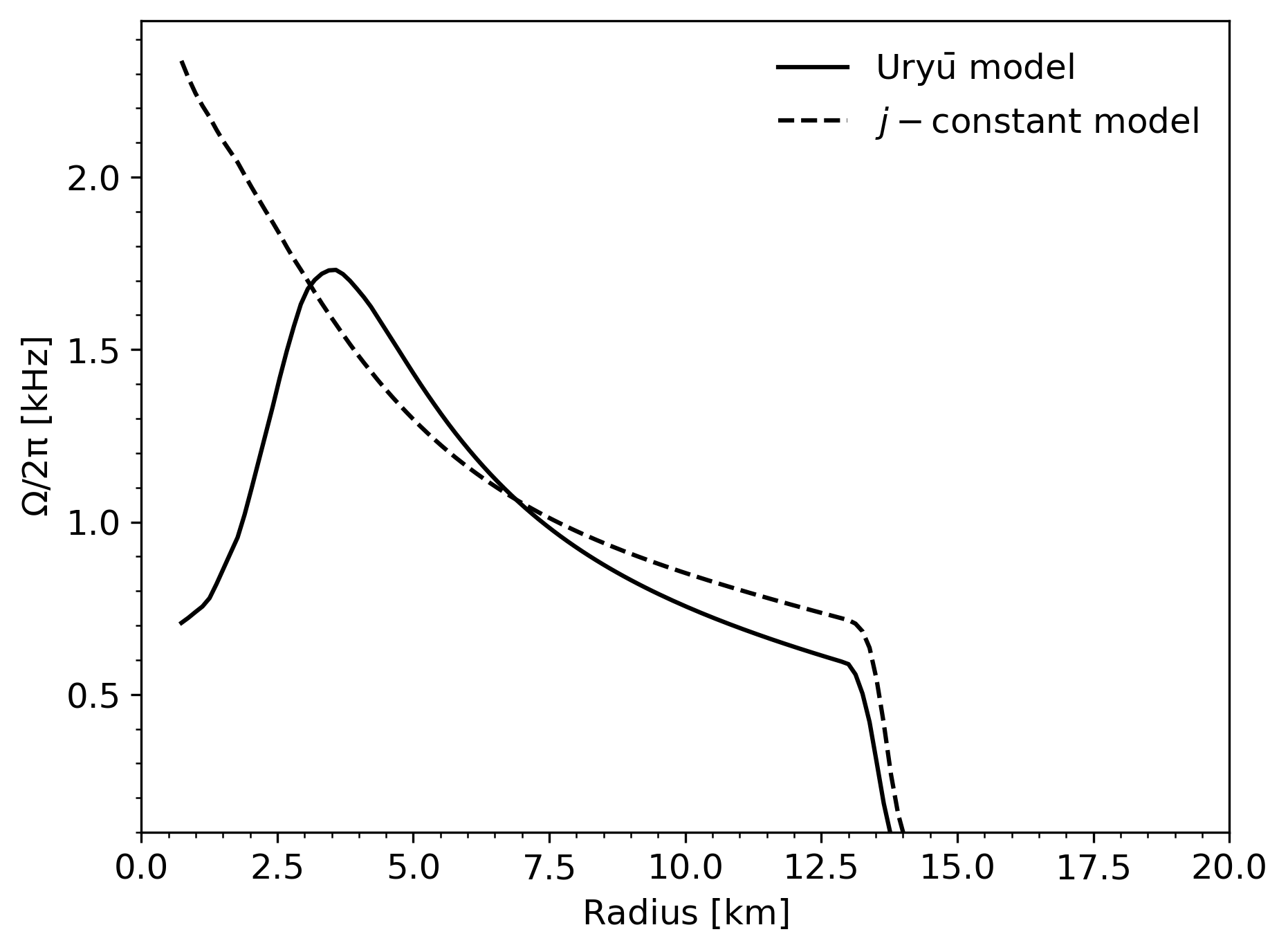}
  \caption{This figure shows the apparent angular velocity profile $\Omega=\nu/\varpi$ for the Ury{\={u}} and $j$-constant models (in the equatorial plane), with $\nu$  the contravariant fluid velocity with respect to an Eulerian observer and $\varpi$  the polar radius. Both tend towards Keplerian profiles at larger radii, but are noticeably different near the rotation axis. We adopt an artificial value for the angular velocity in the surrounding medium, which appears to have little impact on the dynamics of the high density matter.} \label{fig:angular_velocity_initial}
\end{figure}

To construct the initial data, we assume a simple polytropic equation of state,  $p(\rho)=K\rho^\Gamma$  with $\Gamma=2$ and $K=100$. For the evolution of the initial profile, an ideal gas law is used. That is, we have $p=(\Gamma_{\rm{th}}-1)\rho\epsilon,$ with $\Gamma_{\rm{th}}=2$, and $\epsilon$ representing the internal energy. Thermal effects for this type of EOS have been explored in neutron star merger simulations \citep{PhysRevD.82.084043}. To study the effects of the rotation profile we set up two initial models with very similar bulk properties (see Table \ref{tab:models} for the main properties of these models), setting  the ratio of kinetic to gravitational potential energy to a moderate value,  $T/|W|\simeq 0.16$. 
The main difference is in the rotation profile (see Fig.\ref{fig:angular_velocity_initial}). The Ury{\={u}} model features a bell-shaped angular velocity distribution, where the peak of the angular velocity is located at a radius of about 3.5 km. For the $j$-constant  model  the maximum angular velocity is located on the rotation axis. Note that neither of these initial models  truly represents a merger remnant, as the matter distribution is truncated at a finite radius whereas a merger tends to lead to an extended disk. This should not have much impact on local features observed in the high-density region, but may affect the evolution of global dynamics. This can be tested by future work, applying our analysis of the dynamics to actual merger evolution.

\begin{table*}[tb!]
  \centering
  \caption{The main properties of the simulated rotating neutron star models. In the rows we report, from top to bottom, the central rest-mass density $\rho_c$, the central energy density $e_c$, the gravitational mass $M$, the rest mass $M_0$, the proper equatorial radius $R_e$, the total angular momentum $J$, as well as the ratio $J/M^2$, the ratio of rotational kinetic energy and the gravitational binding energy $T/|W|$, the angular velocity at the center $\Omega_c$ and at the equator $\Omega_e$, The Kepler angular velocity $\Omega_{\rm{Kepler}}$, the ratio between the polar and the equatorial coordinate radii $R_p/R_e$. The two models are constructed to represent ``the same star'' and it should be noted that the global quantities (such as mass, angular momentum, and $T / |W|$), all match to around 1\%.} \label{tab:models}
  \begin{tabular}{l|c c c}
    \hline\hline
    Properties & Ury{\={u}} model& $j$-const model& Relative difference \\
    \hline
    $\rho_c [\times 10^{-3}]$ & $1.2801$ & $1.2800 $ & $7.8 \times 10^{-5}$\\
    $e_c [\times 10^{-3}]$ & $1.4438$ & $1.4438$ & $0$ \\
    $M$ & $2.1125$ & $2.1201$ & $3.6 \times 10^{-3}$ \\
    $M_0$ & $2.3031$ & $2.3142$ & $4.8 \times 10^{-3}$ \\
    $R_e$ & $11.506$ & $11.652$ & $1.27 \times 10^{-2}$ \\
    $J$ & $3.4124$ &  $3.4844$ & $2.1 \times 10^{-2}$ \\
    $J / M^2$ & $0.76465$ & $0.77524$ & $1.4 \times 10^{-2}$\\
    $T / |W|$ & $0.15853$ & $0.15857$ & $2.5 \times 10^{-4}$ \\
    $\Omega_c$ & $0.010738$ & $0.035974$ & $2.35$ \\
    $\Omega_e$ & $0.018542$ & $0.021485$ & $0.16$ \\
    $\Omega_{\text{Kepler}}$ & $0.037688$ & $0.037092$ & $0.16$ \\
    $R_p / R_e$ & $0.55$ & $0.515$ & $6.44 \times 10^{-2}$ \\
    \hline\hline
  \end{tabular}
 \end{table*}
 
\subsection{Analysis tools}

Given the two differential rotation laws, we use different quantities to monitor the development of the fundamental instability. First of all, we calculate the quadrupole moment of the matter distribution,
\begin{equation}\label{eq:quadrupole}
  I^{ij} = \int dx^3 \sqrt{\gamma}u^0 \rho x^{i}x^{j},
\end{equation}
in terms of the conserved density $\sqrt{\gamma} u^0 \rho$, where $\gamma$ is the determinant of the three-metric $\gamma_{ij}$ and $u^{\mu}$ is the fluid 4-velocity.  Using the three components of the quadrupole moment in the $x-y$ plane,  we then calculate the
distortion parameters $\eta_{+}$ \citep{2001ApJ...548..919S,2002MNRAS.334L..27S,2003MNRAS.343..619S,2007PhRvD..75d4023B,2010CQGra..27k4104C}, defined as
\begin{equation}\label{eq:eta}
  \eta_{+}(t) = \frac{I^{xx}(t)-I^{yy}(t)}{I^{xx}(0)+I^{yy}(0)}.
\end{equation}
This measure serves as a proxy for the amplitude of global oscillation modes.
To describe the development and saturation of the instability, we also compute the volume-integrated azimuthal density
mode decomposition \citep{2016PhRvD..93b4011E,2015PhRvD..92l1502P,2016CQGra..33x4004E,PhysRevD.100.043014},
\begin{equation}\label{eq:mode}
  C_m(t) = \int dx^3 \sqrt{\gamma}u^0\rho e^{im\phi}\ , 
  \end{equation}
  where $\phi=\mathrm{tan}^{-1}(x/y)$ is the azimuthal angle.

For our simulations, the numerical domain extends to $\sim 300 M_{\odot}$ where the extraction of the gravitational wave signal is plausible. We extract this signal using the Newman-Penrose scalar $\psi_4$ \cite{1962JMP.....3..566N}. This quantity,  calculated by the \emph{Einstein Toolkit} module \emph{WeylScal4}, is decomposed in spin-weighted spherical harmonics of spin-weight $s=-2$ by the  \emph{Multipole} module. The output is the decomposition coefficient $\psi_4^{lm}$, defined as
  \begin{equation}
    \psi_4^{lm}(t,r) = \int {}_{s}Y_{lm}^{*} \psi_4(t,r,\theta,\phi)r^2d\Omega,
  \end{equation}
where $d\Omega$ stands for the differential solid angle, ${}_sY_{lm}^{*}$ the complex conjugate of the spin-weighted spherical harmonics ${}_sY_{lm}$. The gravitational-wave strain $h$ is linked to $\psi_4$ by 
  \begin{equation}
    h = h_{+}-ih_{\times} = \int_{-\infty}^tdt^{\prime} \int_{-\infty}^{t^{\prime}}dt^{\prime \prime}\psi_4\ .
  \end{equation}
 To get the strain, we use the fixed-frequency integration method of \cite{2011CQGra..28s5015R} (see also \citet{2016LRR....19....2B}). We choose a cutoff-frequency of 0.01 in code units, which is smaller than the initial instantaneous frequency of the waves (see Table \ref{tab:models}). To reduce boundary effects, we taper the signal using a Planck window function, with tapering width based on a period corresponding to the cutoff frequency. We also cut off the tapered parts in the final result, following the implementation in the  \emph{PyCactusET} module.
To facilitate a comparison with detector sensitivity curves, we calculate the square root of the power spectrum density (PSD) \cite{2015CQGra..32a5014M}:
  \begin{equation}
    \sqrt{S_h(f)} = 2f^{1/2}|\tilde{h}(f)|,
  \end{equation}
 where $S_h(f)$ is the signal PSD, and $\tilde{h}$ has been calculated as the effective strain 
\begin{equation}
  \widetilde{h}(f) = \sqrt{1/2(\widetilde{h}_{+}^2+\widetilde{h}_\times^2)} \ , 
\end{equation} 
for simplicity.
  
\section{Results and discussion}\label{sec:results}

The stability of axi-stationary fluids has been investigated in detail in the past. Whilst most instability criteria are not relevant for postmerger remnants, one possibility is the existence of a \emph{corotation point}. Loosely this is where the pattern speed of a wave matches the angular velocity. For the particular profiles chosen here, evidence for the presence of corotation point instabilities was provided by \citep{2020arXiv200310198P}. 

All the simulations we discuss represent a star with at least one corotation point. We associate the instabilities that we see for these specific sets of initial data with the existence and properties of the corotation points.
Corotation points have been linked to instabilities in a number of other situations, particularly in disks. The two standard examples are the Papaloizou Pringle instability (PPI) and the Rossby wave instability (RWI). In both cases there is a frame associated with the corotation point with respect to which a wave has negative energy on one side of the corotation point and positive energy on the other side. The interaction across the corotation point leads to a transfer of energy and angular momentum. With the wave reflecting off some boundary there is a positive feedback leading to an instability.
As we are simulating a star and not a disk the detailed theoretical description of the precise type of instability will not carry over to our case. We have the key prerequisites: at least one corotation point, with the reflecting boundary at the inner edge of the disk replaced by the center of the neutron star, and the reflecting boundary at the outer edge of the disk replaced by the surface. However, the geometry is significantly different. In addition, the PPI, in particular, and disk physics, in general, is usually discussed at or near a $j$-constant state. For the Ury{\={u}} case we are (at least initially) far from having constant specific angular momentum, hence the possibility of the existence of two corotation points. As such, we will present analogies to the PPI and RWI cases but will not be able to identify the instability we observe with one rather than the other.

A detailed discussion of the stability of the system will be given in Sec. \ref{sec:stability}. First we will consider the nonlinear development of the instability. The key questions are then how does the difference in the rotational profile influence the development of the instability, and what is its impact on observables like the emitted gravitational waves?

\subsection{Mode analysis}

\begin{figure*}[!ht]
  \centering  
  \subfloat[]{
    \includegraphics[clip,width=0.5\columnwidth]
                        {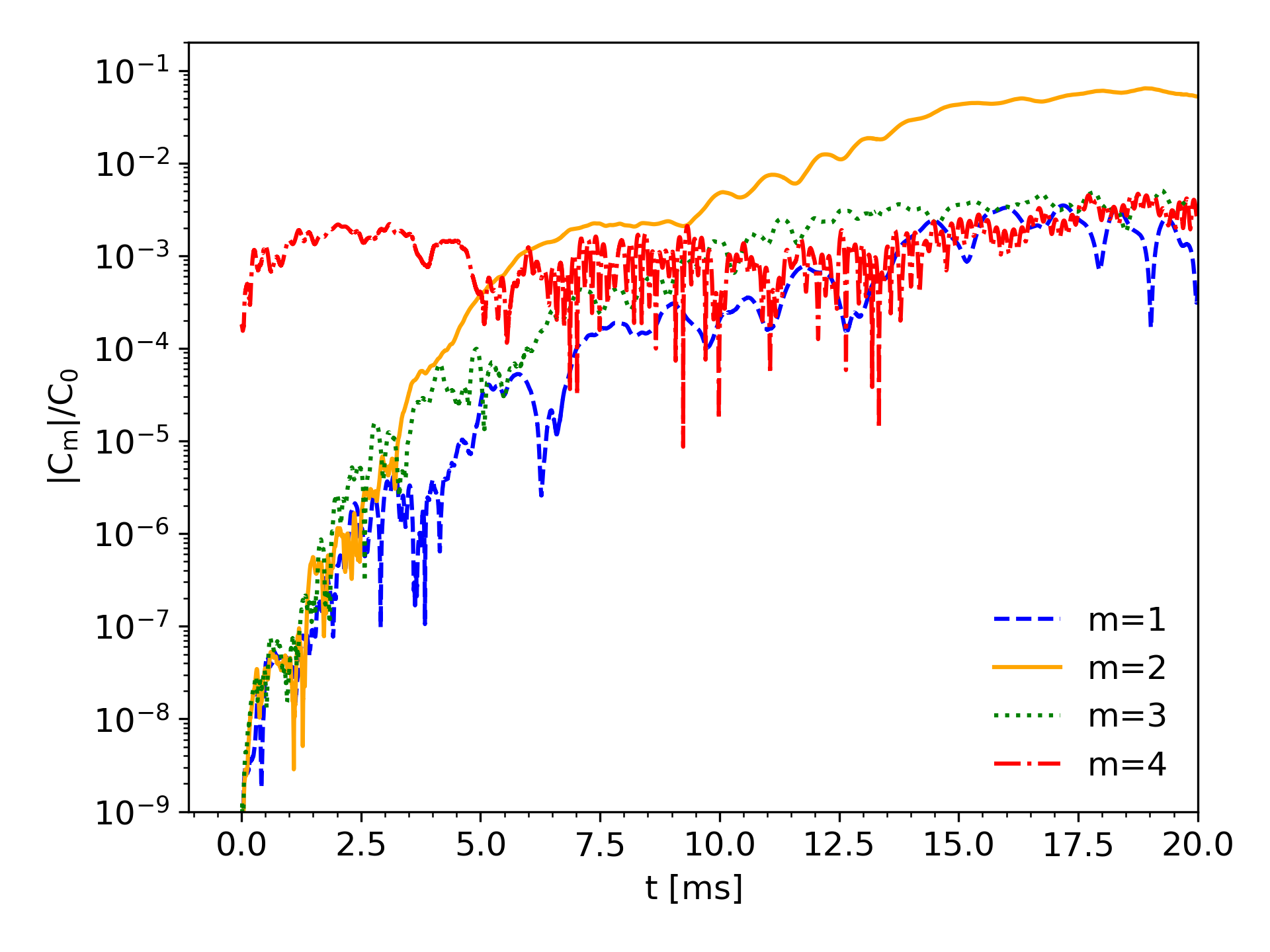}
  }
  \subfloat[]{
    \includegraphics[clip,width=0.5\columnwidth]
                    {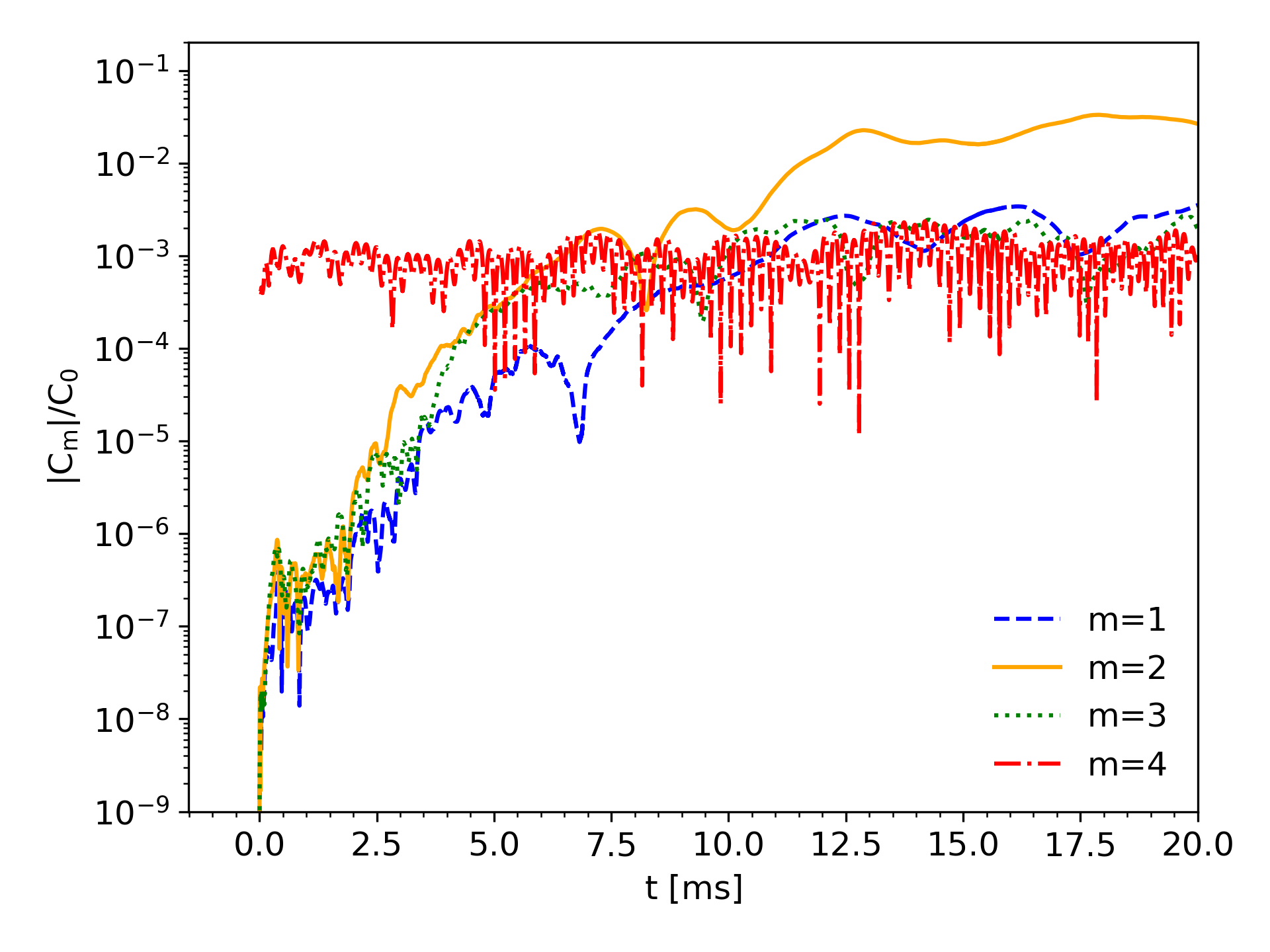}
  }
  \caption{Density mode decomposition for the Ury{\={u}} model (left panel) and the $j$-constant model  (right). We show the normalized magnitude results for modes with azimuthal number in the range $m=1{-}4$. The $m=4$ mode is dominated by noise from the Cartesian grid, leading to its magnitude being stable throughout the simulation. All the other modes grow with the same rate in the beginning. For the Ury{\={u}} model, the $m=2$ mode begins to dominate at around 5 ms. Meanwhile, for the $j$-constant model, the $m=2$ and $m=3$ modes appear to be coupled for the first 10 ms. \label{fig:mode} }
\end{figure*}

In order to understand the global behavior of the instability, we track the nonaxisymmetric modes of matter by extracting the Fourier amplitude of the density variations for the first four azimuthal multipoles,  $m=1{-}4$: see Eq. \eqref{eq:mode}. The time-dependent behavior of the magnitude of $C_m$ illustrates the growth rate of individual modes and the nonlinear coupling associated with an instability \citep{2007PhRvD..75d4023B} and is shown for both the Ury{\={u}} model and the $j$-constant model in Fig \ref{fig:mode}.

As the initial data for the rotating models is close to an unperturbed axisymmetric equilibrium, at the beginning of the simulation the magnitude of each mode with azimuthal number $m=1{-}3$ is essentially zero. However, the Cartesian grid naturally introduces an $m=4$ mode from the numerical discretization error. The dynamics of this $m=4$ mode is similar to previous findings in the literature \citep{2007PhRvD..75d4023B,2010CQGra..27k4104C,2015PhRvD..92l1502P}. For this reason we consider the $m=4$ mode to be ``grid noise'' in these simulations.

The other nonaxisymmetric modes ($m=1{-}3$) rapidly grow together from the initial numerical noise. For the Ury{\={u}} model, the $m=2$ mode begins to play a major role after about 5 ms. Meanwhile, for the $j$-constant model, the $m=2$ and $m=3$ modes notably grow at a similar rate for the first 10 ms. At late times the $m=2$ azimuthal mode dominates over all other modes in both cases. For the $j$-constant model the contribution of the $m=1$ and $m=3$ modes is just above the $m=4$ grid noise, but is--at the global level--marginal.

\subsection{Gravitational waves}

\begin{figure*}[!ht]
  \centering  
  
  \subfloat[]{
    \includegraphics[clip,width=0.5\columnwidth]
                        {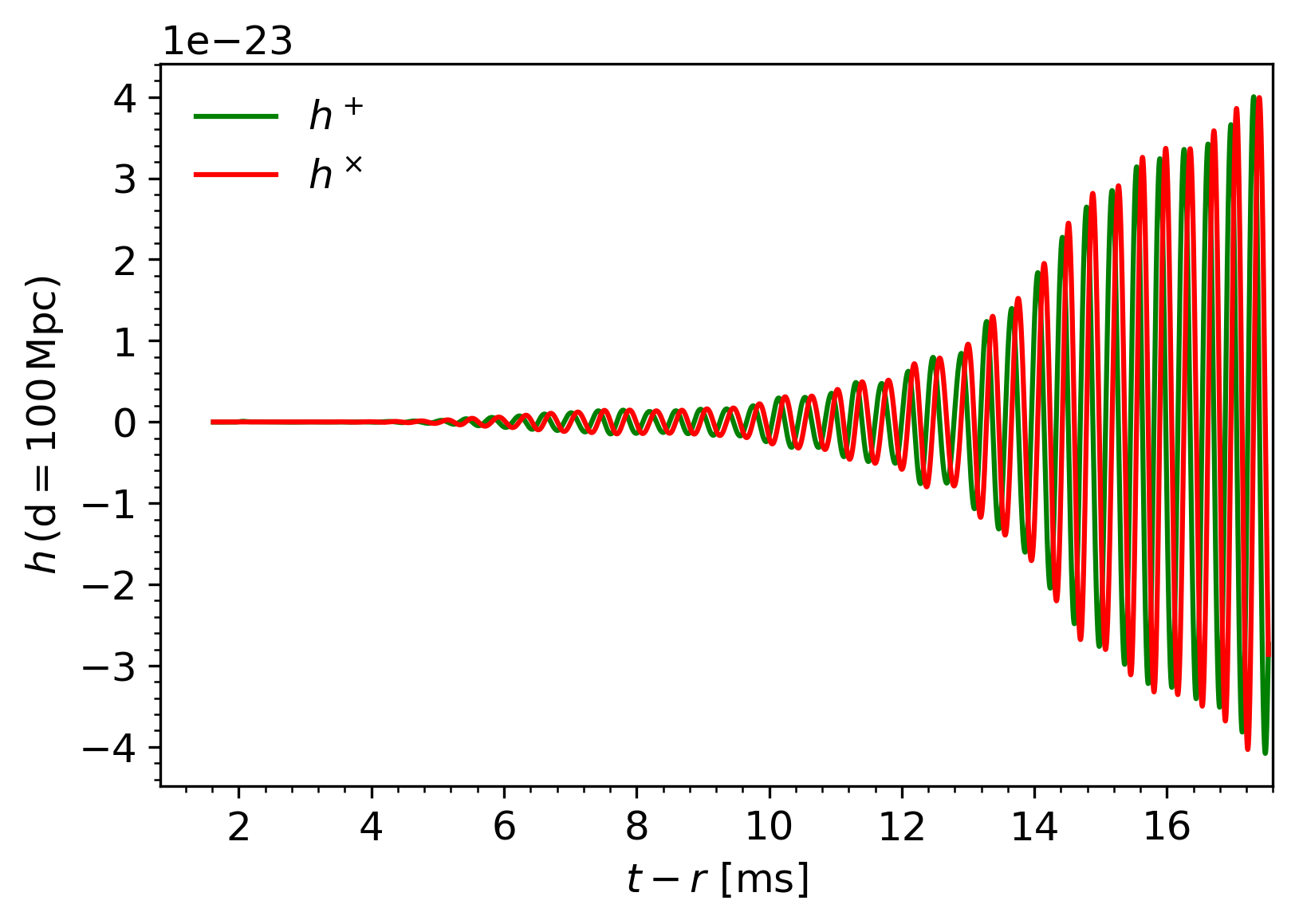}
  }
  \subfloat[]{
    \includegraphics[clip,width=0.5\columnwidth]
                    {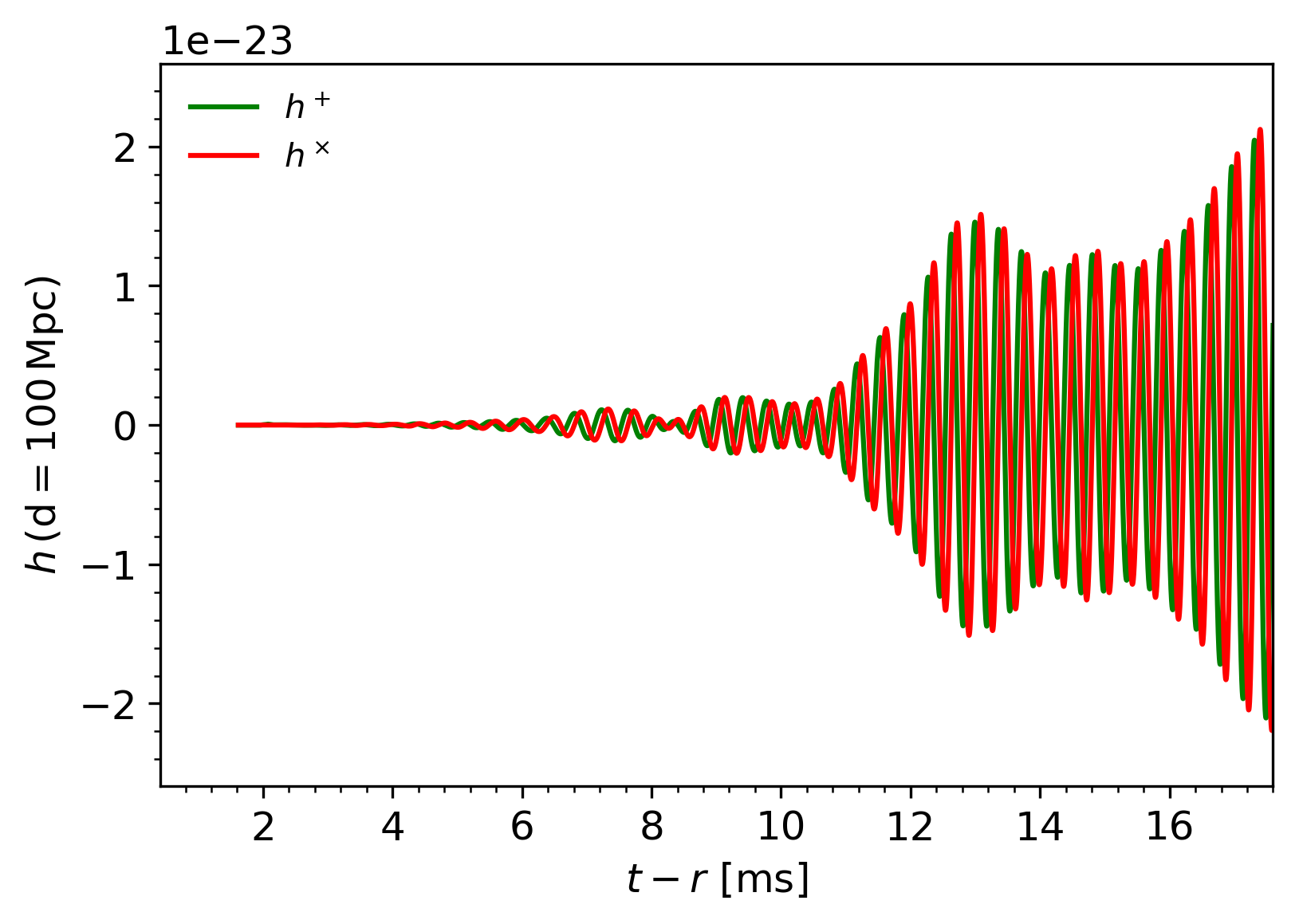}
  }
  \caption{The gravitational-wave strains, $h_{+}$ and $h_{\times}$,  as a function of the retarded time $t-r$ for the Ury{\={u}} model (left) and the $j$-constant model (right). The observer is located on the $z$ axis at a fiducial distance of 100~Mpc.\label{fig:gw_strain}}
\end{figure*}

Dynamically unstable rotating stars, as they deform to nonaxisymmetric configurations, may be relevant sources of quasiperiodic gravitational waves (see e.g.\ \citet{2002MNRAS.334L..27S}). The gravitational-wave signature depends on the specific evolution of the leading quadrupole mode and the extent to which other modes suppress its growth. \citet{2007PhRvD..75d4023B} find that generic nonlinear mode-coupling effects appear during the development of the instability, and these can severely limit the persistence of any bar-mode deformation. Despite the initial data used here having a $T/|W|$ well below that needed for the classical bar mode instability, we note the growth of low-$m$ modes in a similar fashion to studies of bar modes such as \citet{2007PhRvD..75d4023B}. Figure \ref{fig:gw_strain} shows the  gravitational-wave strain as viewed by an observer located on the $z$-axis at a distance of 100~Mpc. This figure should be viewed alongside Fig. \ref{fig:mode}, as there is an association between gravitational-wave pattern and oscillation modes, which can also be found in previous work (e.g.\ \cite{2010CQGra..27k4104C}).

\begin{figure*}[!ht]
  \centering  
  
  \subfloat[]{
    \includegraphics[clip,width=0.5\columnwidth]
                        {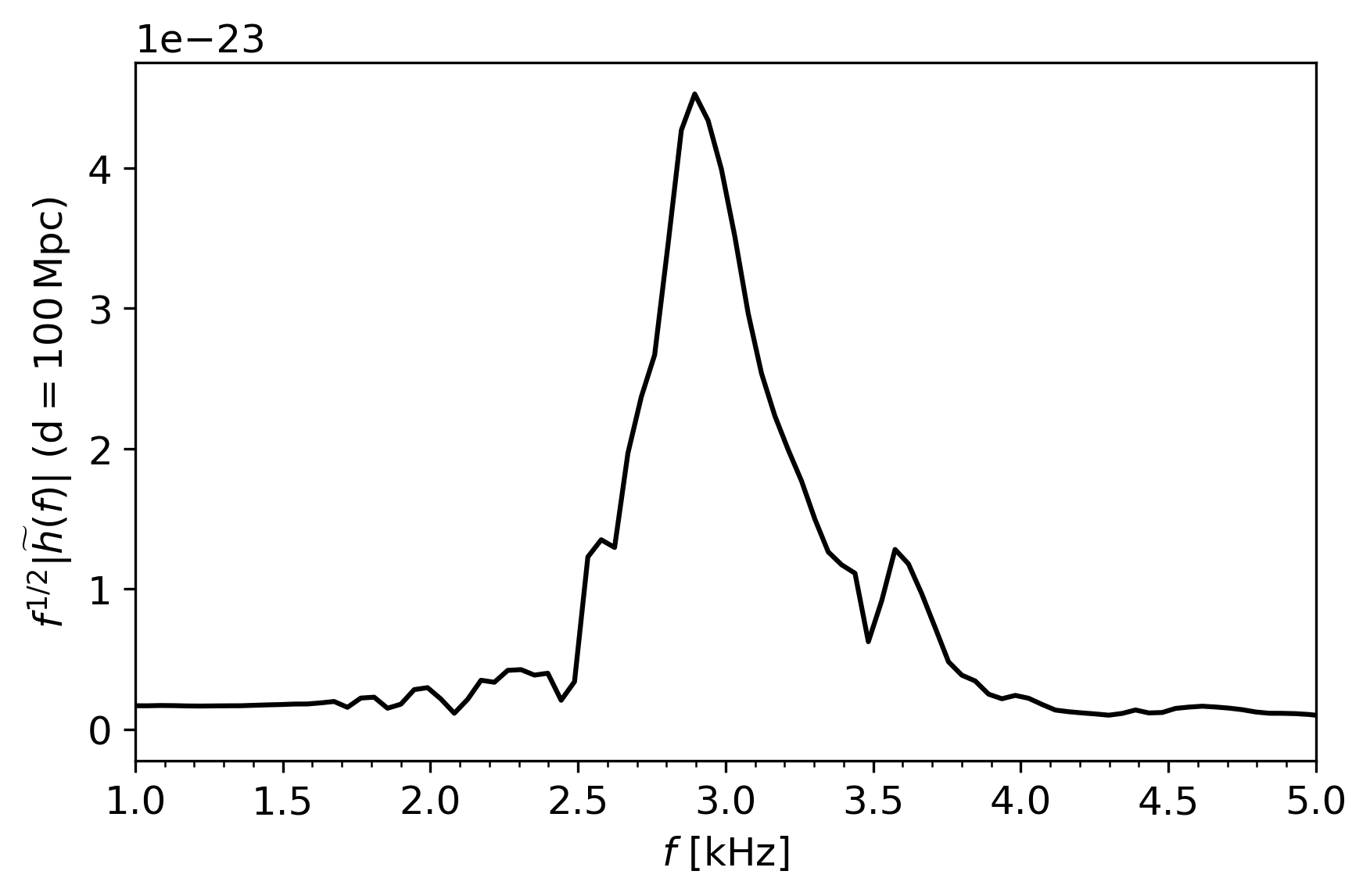}
  }
  \subfloat[]{
    \includegraphics[clip,width=0.5\columnwidth]
                    {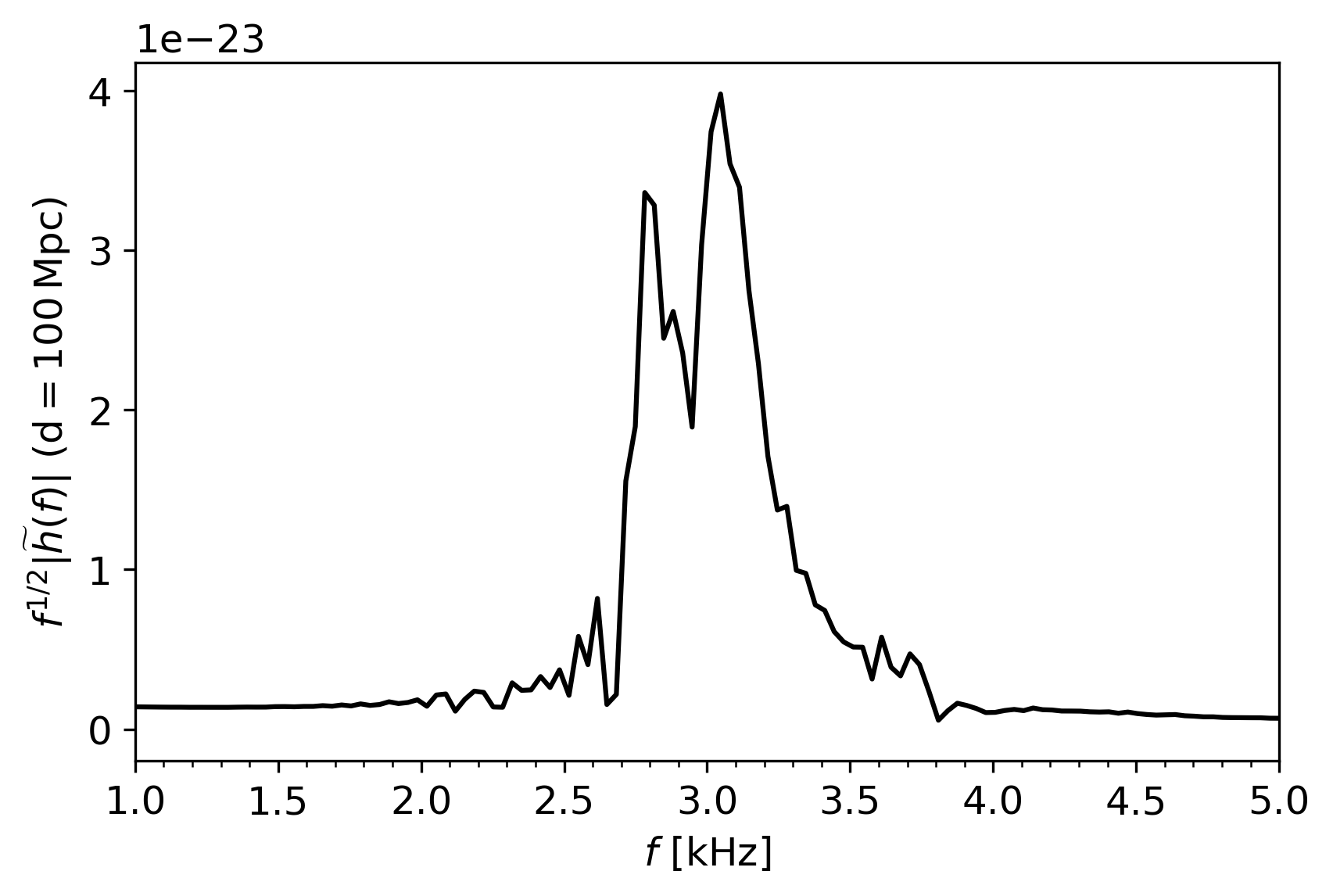}
  }
  \caption{The amplitude spectral density of gravitational waves, $f^{1/2}|\widetilde{h}(f)|$, observed at 100~Mpc for the Ury{\={u}} model (left) and the $j$-constant model (right). The effective strain is defined as $\widetilde{h}(f) = \sqrt{1/2(\widetilde{h}_{+}^2+\widetilde{h}_\times^2)}$. Both  profiles show a dominant peak at nearly the same frequency. For the Ury{\={u}} model, the peak is at 2.9 kHz, while the peak is located at at 3.0 kHz for the $j$-constant model. \label{fig:gw_psd}}
\end{figure*}

The main features are the same in both models. As the instability grows, a long-lived gravitational wave of modest amplitude is generated in both polarizations. As shown in the previous section, the instability is dominated by an $m=2$ mode throughout the simulation, which in turn dominates the gravitational wave emission. We distinguish some modulation of the signal on long time scales in the $j$-constant model. By comparing the gravitational wave emission in Fig. \ref{fig:gw_strain}(b) to the detailed mode growth in Fig. \ref{fig:mode}(b) it could be argued that this modulation is linked to the $m=2$ mode behavior. However, as all modes appear to have saturated by the end of simulation for both models, this suggests that the gravitational wave emission will remain largely unchanged for both models, and so the two models will be difficult to distinguish.

The power spectra associated with the gravitational-wave signals are shown in
Fig. \ref{fig:gw_psd},  comparing the Ury{\={u}} (left) and $j$-contant (right) models. The main peak, located at around 3~kHz, is very similar for the two  rotation profiles, and is in accordance with the postmerger gravitational wave frequency found in merger simulations (see e.g. \cite{2015PhRvL.115i1101B,2017PhRvD..95b4029D,2017PhRvD..95d4045D})

\subsection{Dynamics of the instability}\label{sec:dynamics}

\begin{figure*}[!ht]
  \centering  
  \includegraphics[clip,width=1.0\textwidth]{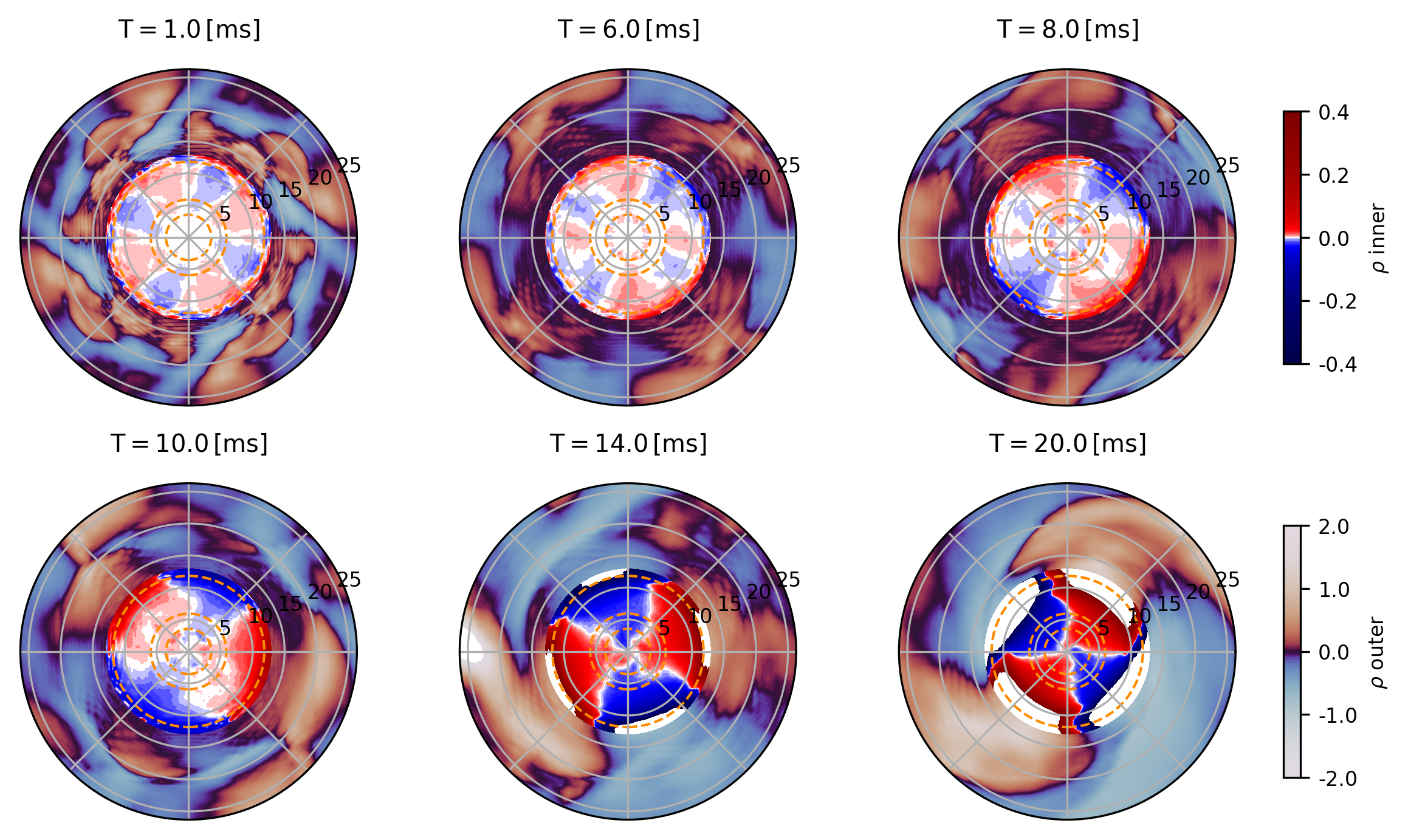}
  \caption{Snapshots of the density variation $\frac{\rho -\langle\rho\rangle}{\langle\rho\rangle}$ in the equatorial plane for the Ury{\={u}} model. The density average $\langle\rho\rangle$ is taken over a circular ring of radius $\varpi = \sqrt{x^2+y^2}$. Two semilog color bars are used to visualize the inner part of the simulation domain i.e. within the neutron star profile and the surrounding low-density medium. The scale of the color-bar is set to $\pm 0.4$ for the inner part, and $\pm 2$ for the outer part. The dotted circle indicates the boundary of the refinement levels (the real grid boundary is square, here we use circle instead for better visualization). The data illustrate how the initial variation  (effectively representing an $m=4$ mode) associated with the Cartesian grid, gives way to the development of an unstable $m=2$ mode. At $t=6$ ms, as the outer $m=2$ mode establishes itself, the inner part of the remnant (inside the innermost refinement level, approximately 3.5 km) develops a local perturbation, initially with a high azimuthal number m ($t=6{-}8$ ms). The \emph{inner} perturbation evolves into a $m=2$ oscillation which rotate relatively faster than the outer $m=2$ mode. By the end of simulation, at $t=20$ ms, the \emph{inner} oscillation almost synchronizes with the outer mode. \label{fig:uryu_density_reflux}}
\end{figure*}

\begin{figure*}[!ht]
  \centering
  \includegraphics[clip,width=1.0\textwidth]{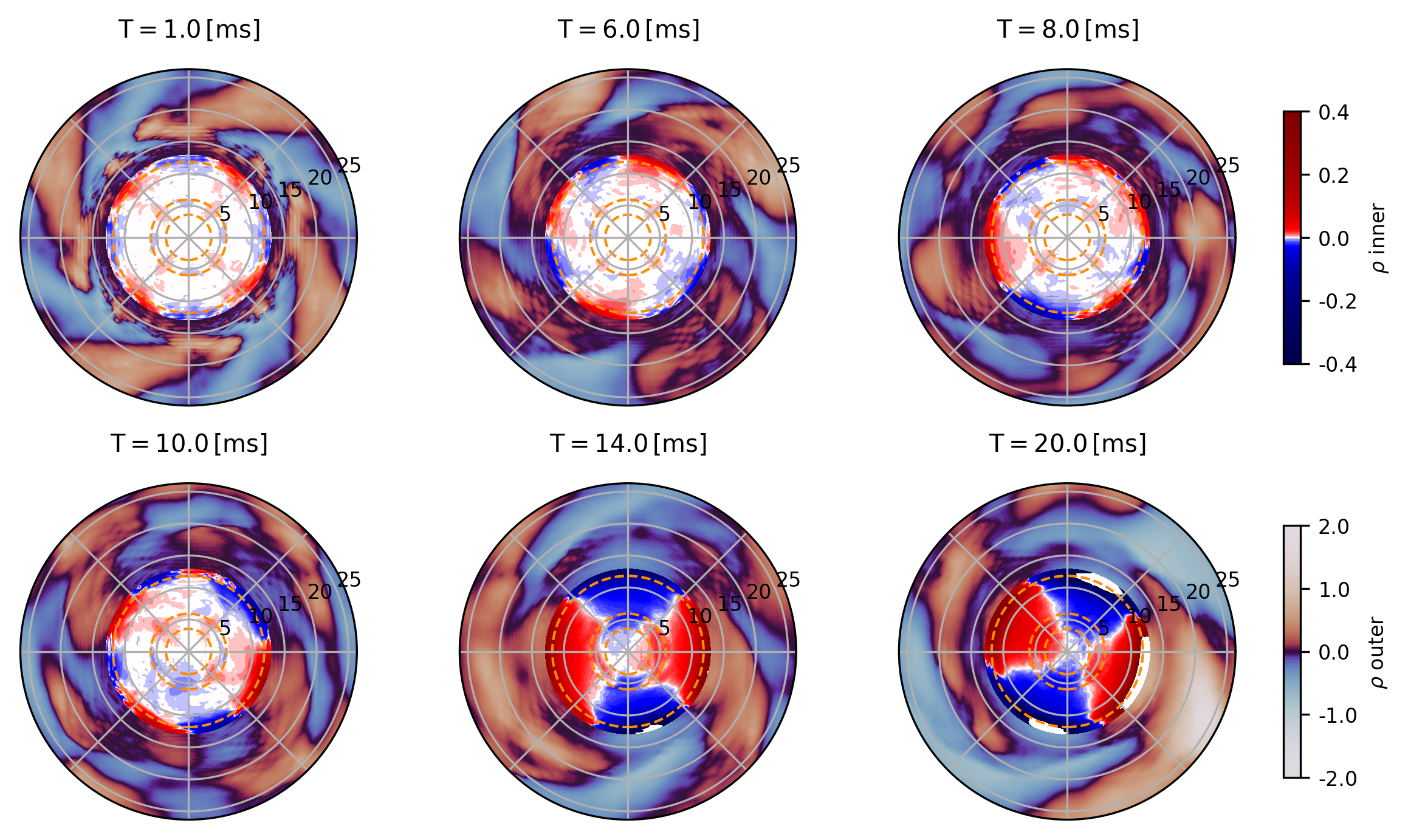}
  \caption{ Same as Fig.\ref{fig:uryu_density_reflux}, but for the $j$-constant rotation law. Compared to the Ury{\={u}} model, no fast-rotating inner part is evident in the contours. Before the time $t=14$ ms, the $m=2$ and $m=3$ modes are both apparent in the contour plots.\label{fig:jconst_density_reflux}}
\end{figure*}

We now consider the dynamics associated with the mode instability, comparing and contrasting the behavior for the Ury{\={u}} and $j$-constant models. In Figs. \ref{fig:uryu_density_reflux} and \ref{fig:jconst_density_reflux}, we show the snapshots of the normalized density deviation from the  averaged value, $(\rho-\langle\rho\rangle )/\langle\rho\rangle$, where $\langle \ldots\rangle$ represents an angular average at a specific radius in the equatorial plane \cite{2020MNRAS.493L.138S}. The computational domain is divided into two parts, approximately representing the neutron star interior and the surrounding low-density medium (the atmosphere). Separate color maps are used to visualize the density variation inside, and outside of, the neutron star. 

In the case of the Ury{\={u}} rotation law, the initial density variation originates from the grid discretization. This leads to a quasistationary $m=4$ mode pattern, evident in the first ($t=1$ ms) snapshot in Fig. \ref{fig:uryu_density_reflux}. At about $t=6$ ms, an $m=2$ mode becomes apparent in the outer part of the remnant for the Ury{\={u}} model, while a high azimuthal number ($m \sim 8$) oscillation appears in the inner region. The azimuthal number of the oscillation reduces to $m=4$ at around $t=8$ ms, and is then replaced by an $m=2$ oscillation at $t=10$ ms (hereafter, we will refer to this oscillation pattern as the \emph{inner local} instability or \emph{inner local} oscillation). As we can see from the panels that represent $t = 10{-}14$ ms, the \emph{inner local} instability in the center rotates faster than the $m=2$ mode in the rest of the remnant. At the end of the simulation (at $t=20$ ms), the \emph{inner local} oscillation in the center has almost synchronized with the rest of the remnant.

In contrast,  the corresponding results for the $j$-constant model, shown in Fig.\ref{fig:jconst_density_reflux}, show two major differences to the Ury{\={u}} model. First of all, we do not observe a distinct local oscillation close to the center. In fact, at $t=6$ ms, the density deviation close to the centre is minimal. Instead, a combination of $m=2$ and $m=3$ modes develops in the outer part of the neutron star. The $m=2$ and $m=3$ modes then coexist for a while, as is evident in the panels that represent $t=6{-}10$ ms. Eventually, the $m=2$ mode takes over and dominates over all the other modes until the end of the simulation. The dynamics of the density deviation is in agreement with the mode analysis results from the previous section.  

\subsection{Rotation profile evolution and corotation radius}

\begin{figure*}[!ht]
  \centering  
  \subfloat[]{
    \includegraphics[clip,width=0.5\columnwidth]
                        {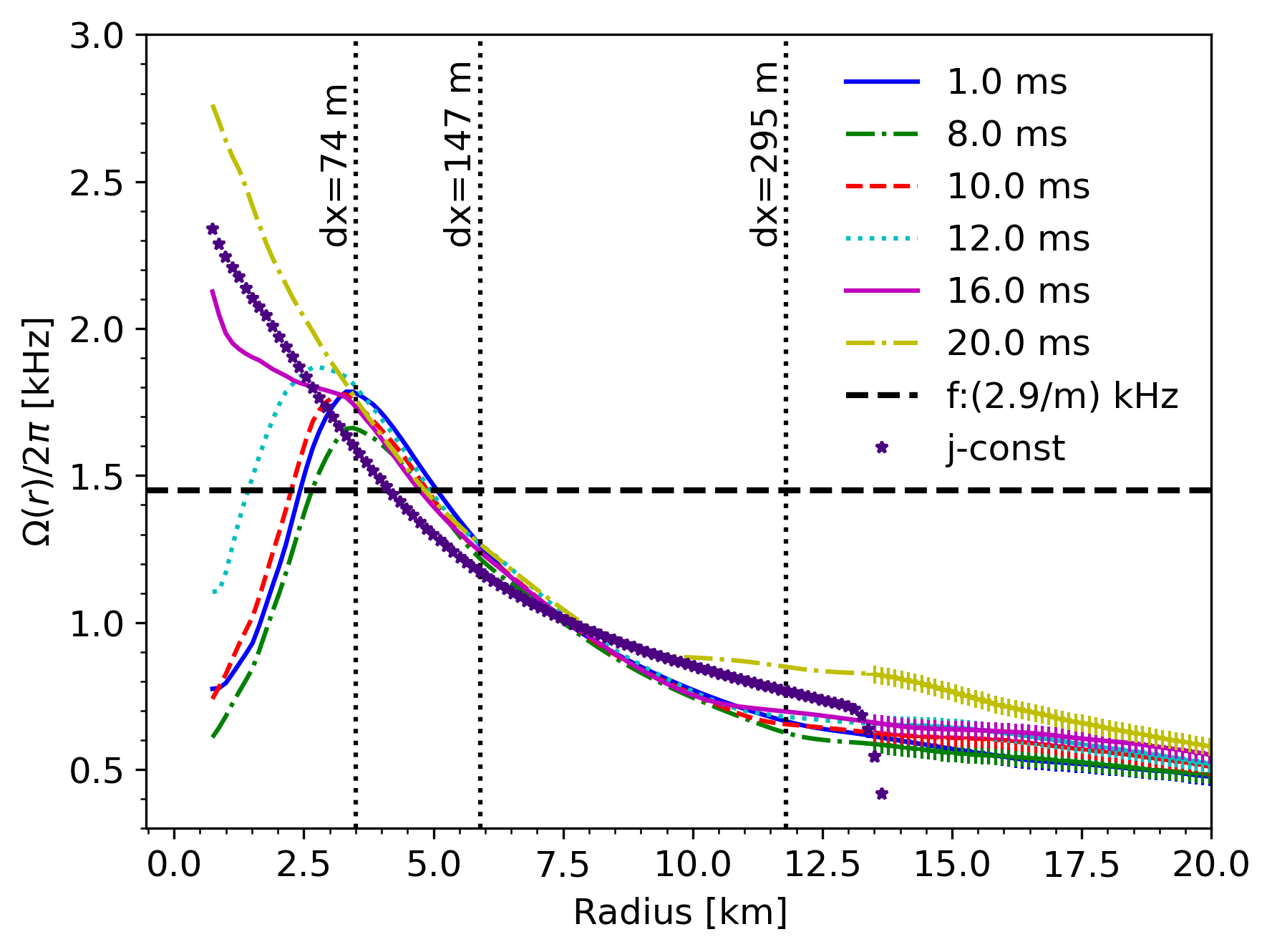}
  }
  \subfloat[]{
    \includegraphics[clip,width=0.5\columnwidth]
                    {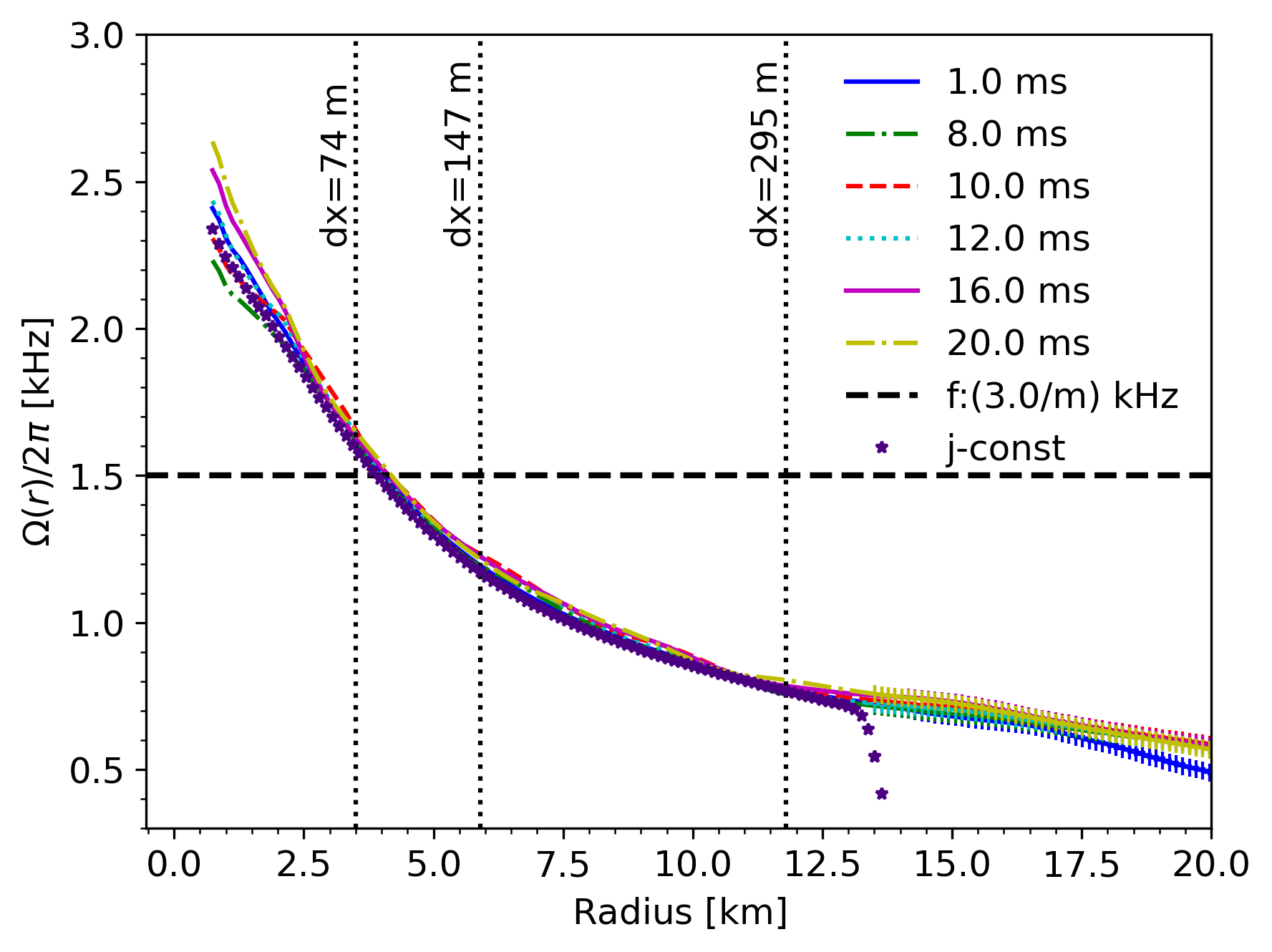}
                    }
  \caption{Time evolution of the angular velocity profile for the Ury{\={u}} model (left) and the $j$-constant model (right). The initial angular velocity of the $j$-constant model is added in each plot for comparison. The vertical lines represent the refinement boundary with the grid spacing indicated.  The horizontal line shows the pattern frequency $f=\sigma_2/2$, where $\sigma_2$ is the gravitational-wave frequency, inferred from Fig.\ref{fig:gw_psd}. The $+$ symbol represents the region in the atmosphere. Matter from the remnant rapidly fills the adjacent atmosphere, producing a smooth angular velocity profile within 1 ms. For the Ury{\={u}} model, the peak of the angular velocity drifts inwards after about $8$ ms. The inner part of the angular velocity profile increases with time, producing a profile similar to that of the $j$-constant model. For the $j$-constant model, the angular velocity profile does not change much within the first $20$ ms of simulation, although the angular velocity at the center increases slightly. \label{fig:angular_velocity_evolution}}
\end{figure*}

Given a specific oscillation mode, we can associate the frequency to the angular velocity of the rotating profile, and then define the corotation point as the position where the mode's pattern speed matches the bulk angular velocity. It has been suggested (see, e.g., \citep{2005ApJ...618L..37W,2010CQGra..27k4104C,2020arXiv200310198P}) that a low $T/|W|$ instability sets in when such corotation points of the unstable modes exist. For the $j$-constant rotation law, only one corotation point can possibly exist for each mode. When we consider the Ury{\={u}} model, however, a given mode may exhibit two distinct corotation points due to the bell-shaped feature of the angular velocity profile (see Fig. \ref{fig:angular_velocity_initial} and the discussion in \cite{2020arXiv200310198P}). 
  
In Fig. \ref{fig:angular_velocity_evolution}, we present the time evolution of the angular velocity profile for both models. The peak frequency of the $l=m=2$ gravitational wave signal is also plotted to infer the mode's corotation radius. For the Ury{\={u}} model, the peak of the angular velocity drifts inward at around $t=8$ ms. The angular velocity of the inner part of the remnant increases with time (see the lines corresponding to $t=10{-}20$ ms), leading to a final profile similar to that of the $j$-constant model. Also from the value of the peak gravitational wave frequency we infer that the pattern speed of the $m=2$ mode is $f\approx 1.45$ kHz. The associated outer corotation radius is at around 5 km, and this outer corotation radius does not change with time. The inner corotation point is however difficult to define as the angular velocity changes with time for the inner region. For the $j$-constant model, the angular velocity profile remains roughly the same during the first 20 ms of simulation. The corotation radius associated with the $m=2$ mode locates at around 4 km, and does not change with time. 

The additional corotation point within the Ury{\={u}} model makes the mode analysis more complicated, leading us to consider the relevance of each corotation point. Will there, for example, be an independent mode pattern associated with the inner corotation point? 

\begin{figure*}[!ht]
  \centering  
  \subfloat[]{
    \includegraphics[clip,width=0.5\textwidth]
                        {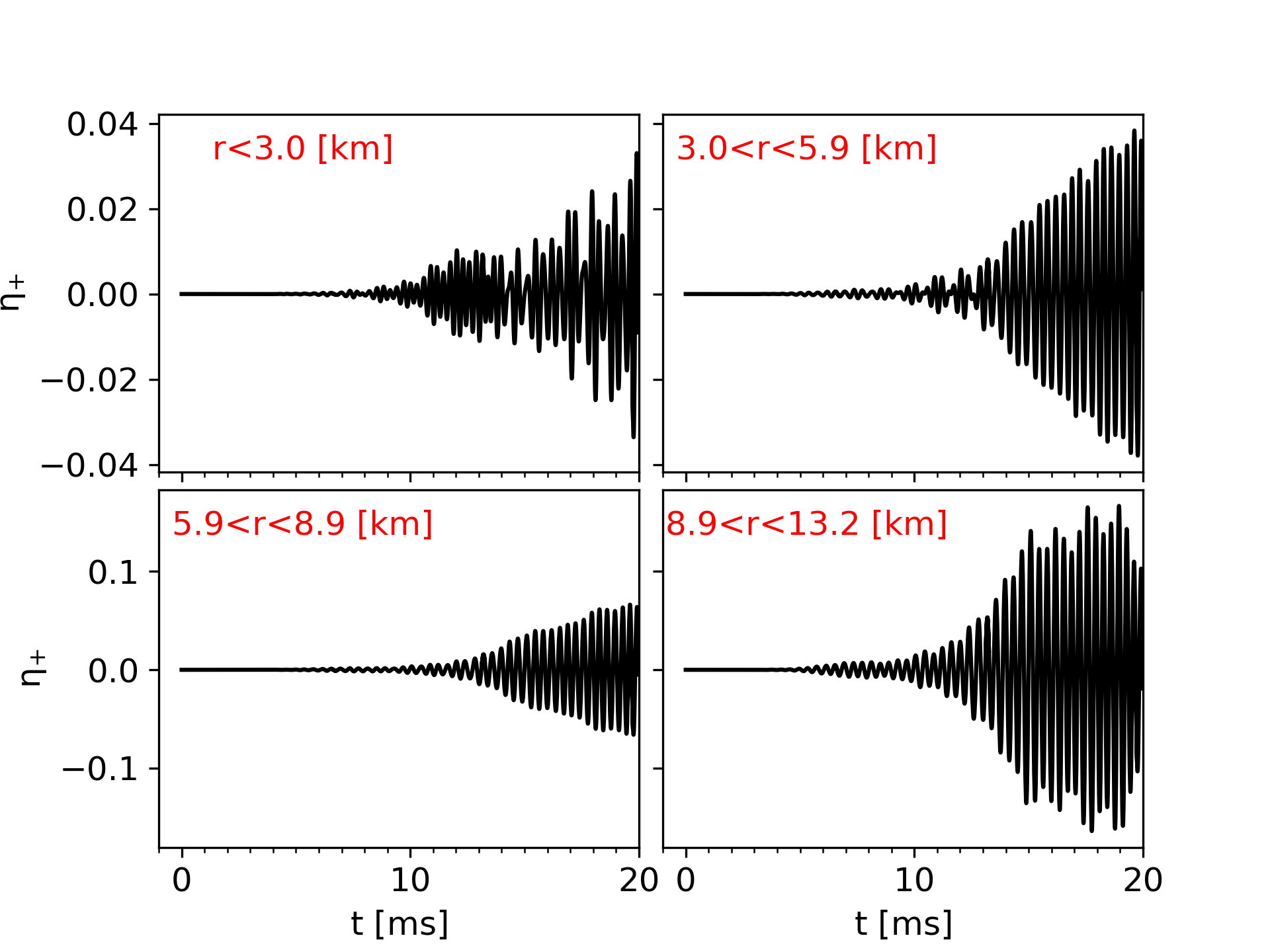}
  }
  \subfloat[]{
    \includegraphics[clip,width=0.5\textwidth]
                    {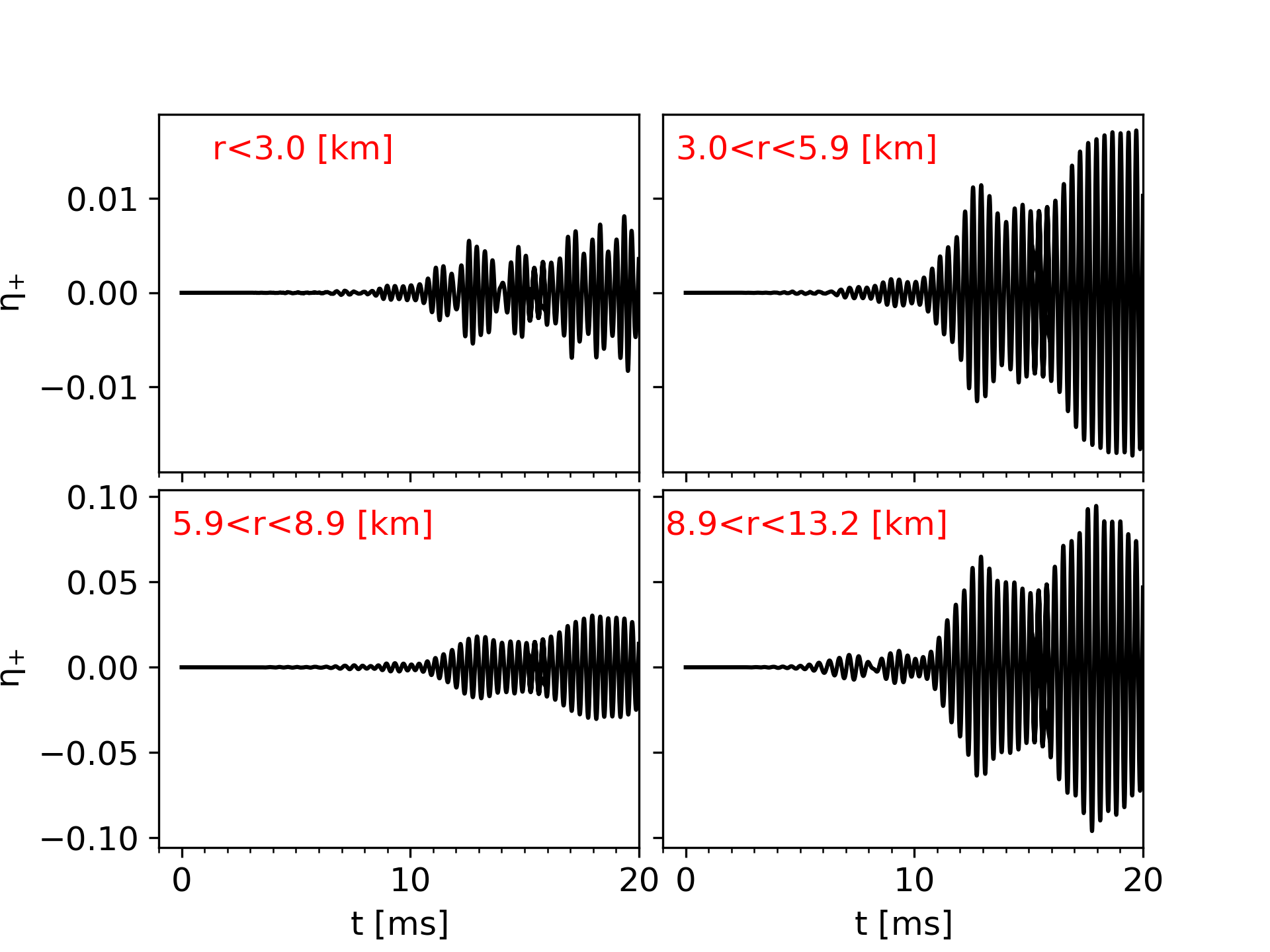}
  }
  \caption{The time evolution of the distortion parameter $\eta_{+}$ within four radial regions for the simulation of the Ury{\={u}} (left panel) and $j$-constant (right) model. \label{fig:eta_plus_ts}}
\end{figure*}

\begin{figure*}[!ht]
  \centering  
  \subfloat[]{
    \includegraphics[clip,width=0.5\textwidth]
                        {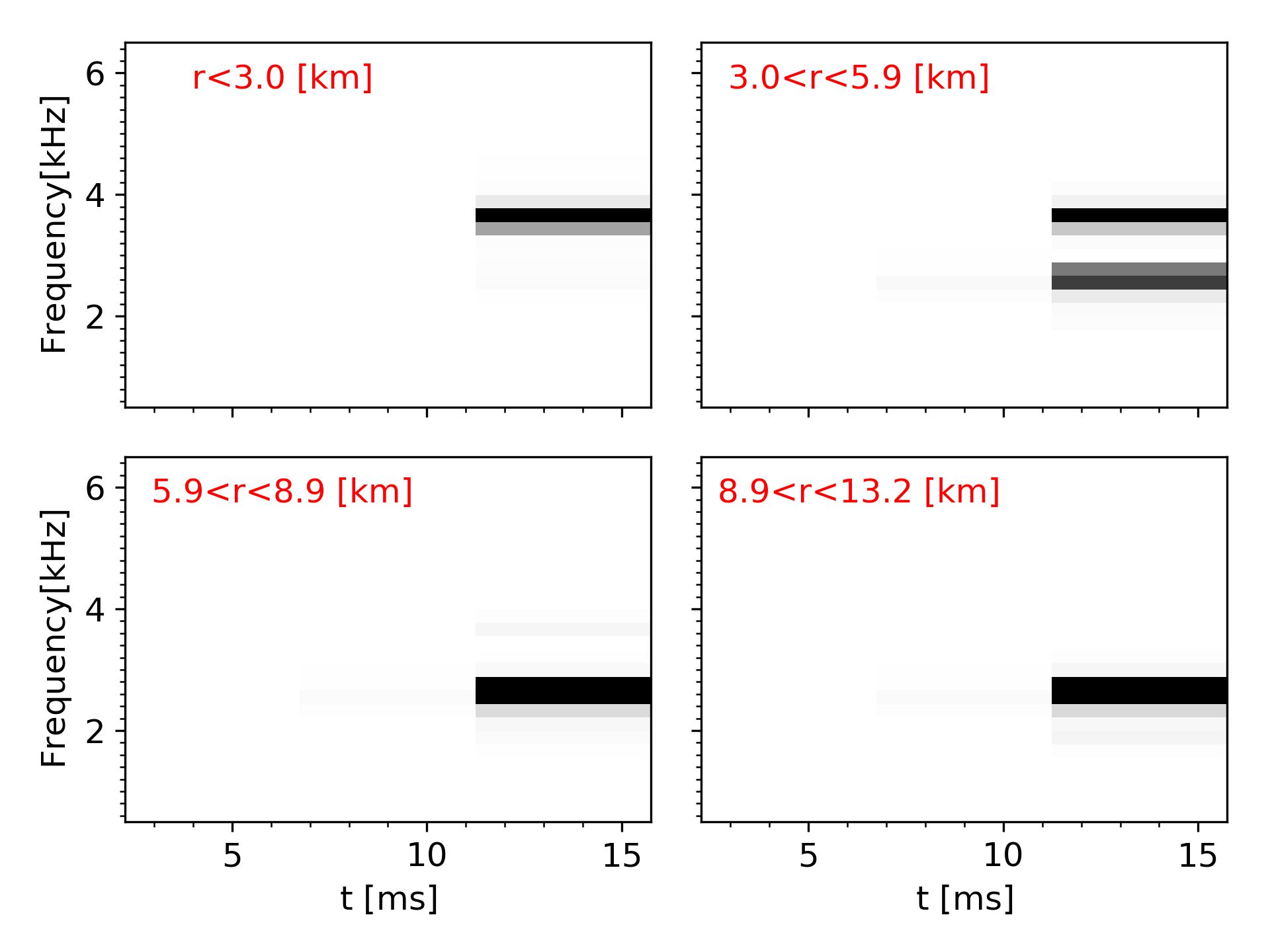}
  }
  \subfloat[]{
    \includegraphics[clip,width=0.5\textwidth]
                    {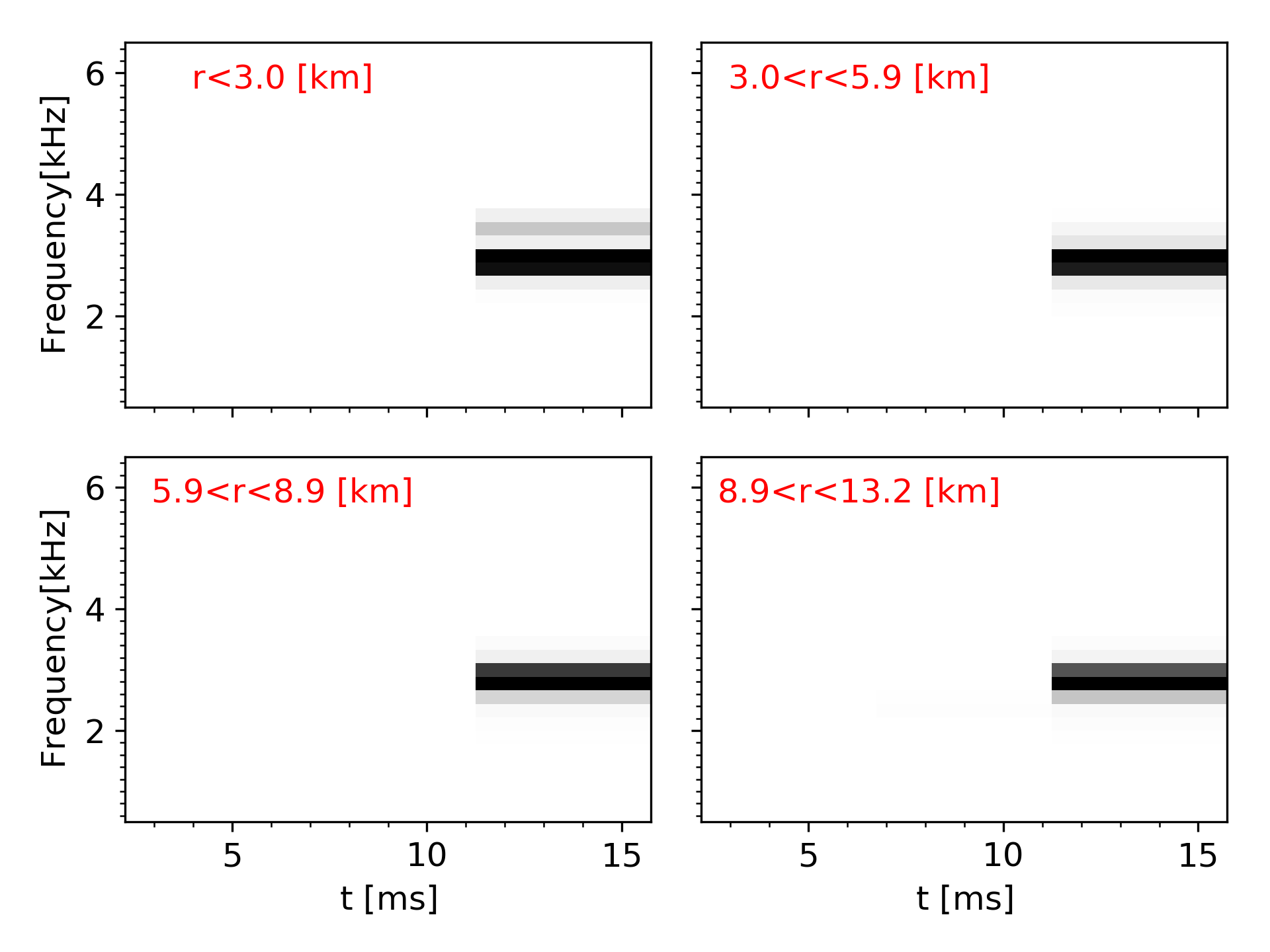}
  }
  \caption{Power spectrum of the Fourier transform of the distortion parameter $\eta_{+}$ within four radial regions for the simulation of the Ury{\={u}} (left panel) and $j$-constant (right) model. The time window adopted for the Fourier transformation is 4.5 ms. The inner region in the Ury{\={u}} model features a separate oscillation mode with a frequency close to 3.6 kHz. The oscillation frequency of the distortion parameter in the rest of the remnant is around 2.8 kHz, in accordance with the gravitational-wave frequency. For the $j$-constant model, the oscillation frequency throughout the remnant is around 2.9 kHz. \label{fig:eta_plus_spec}}
\end{figure*}

Based on the results of the density deviation plots for the Ury{\={u}} model, we find that the oscillation pattern close to the center rotates faster throughout the simulation. To analyze this \emph{inner local} oscillation, we calculate the distortion parameter $\eta_{+}$ [see Eq. \eqref{eq:eta}] for four adjacent radial regions: 1. $r<3\,\rm{[km]}$, 2. $3.0<r<5.9\,\rm{[km]}$, 3. $5.9<r<8.9\,\rm{[km]}$, and 4. $8.9<r<13.2\,\rm{[km]}$. The results are shown in Fig. \ref{fig:eta_plus_ts}. Note that the inner local oscillation, inferred from Fig. \ref{fig:uryu_density_reflux}, is located within a radius of 5 km. The first region then represents the inner core. The second region bridges the core with the outer envelope. The third and fourth regions represent the envelope of the remnant.  We calculate the fast Fourier Transform (FFT) of the time series of the distortion parameter $\eta_{+}$ inside each region. We divide the time series into windows with a width of $4.5\,\rm{ms}$ to perform the FFT calculation. The power spectra are shown in Fig. \ref{fig:eta_plus_spec}. For the Ury{\={u}} model, the inner region reveals a clear oscillation mode at a frequency of 3.6 kHz during the time interval $11.25\,\rm{ms}{-}15.75\,\rm{ms}$. The third and fourth regions reveal a mode at a frequency of 2.8 kHz. The second region shows two mode frequencies. One corresponds to the inner \emph{local} oscillation, the other corresponds to the \emph{outer} mode in the rest of the remnant. For the $j$-constant model, a single mode with frequency at around 2.9 kHz covers all of the regions.

\subsection{The \emph{inner local} instability}\label{sec:inner}

\begin{figure*}[!ht]
  \centering  
  \includegraphics[clip,width=1.0\columnwidth]
                  {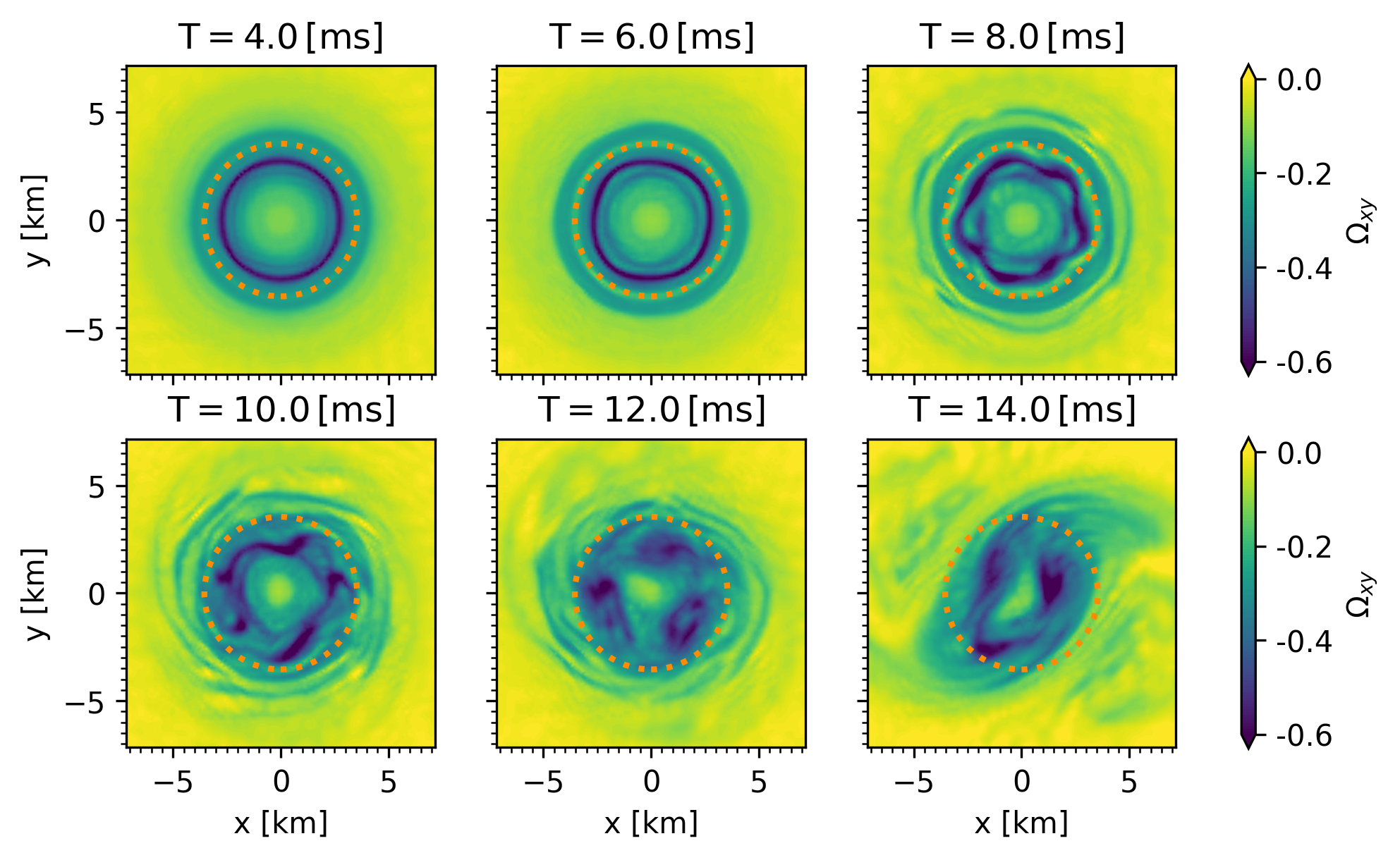}
  \caption{The time snapshots of the equatorial vorticity $\Omega_{xy}$ contour for the Ury{\={u}} model. The dotted circle indicates the peak of the angular velocity. From the time around 6 ms, the instability inside the peak region drives nonasymmetric perturbation which rotates with the flow. The perturbation becomes unstable at around 8 ms and forms vortex islands. Those vortices further merge into three large vortices at around 12 ms and eventually, merge into a big vortex in the center. \label{fig:uryu_vorticity}}
\end{figure*}

\begin{figure*}[!ht]
  \centering  
  \includegraphics[clip,width=1.0\columnwidth]
                  {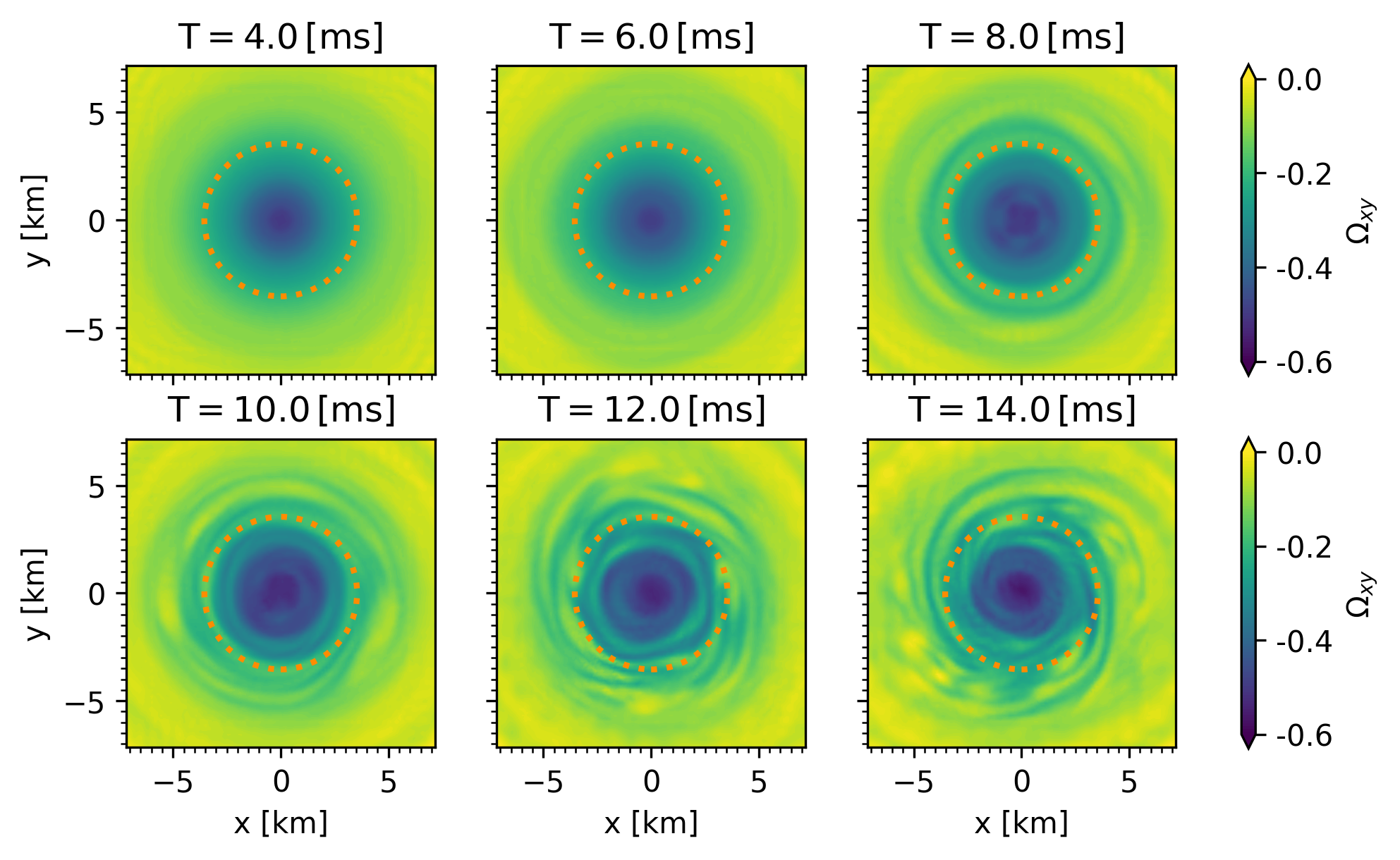}
  \caption{Same as Fig. \ref{fig:uryu_vorticity}, but for the $j$-constant model. The dotted circle is located at the same radius as in Fig. \ref{fig:uryu_vorticity}. Compared to the Ury{\={u}} model, a persistent rotating spiral pattern develops around the center. \label{fig:jconst_vorticity}}
\end{figure*}

In this section, we discuss the nature of the inner fast-rotating \emph{local} oscillation for the Ury{\={u}} model. To minimize the grid effects, we conduct two further simulations for both models without the innermost refinement level, which has the boundary set at around 3.5 km (close to the peak of the angular velocity profile in the Ury{\={u}} case). The innermost refinement boundary for these  new simulations extends to a radius of about 6 km, covering the entire peak region of the angular velocity for the Ury{\={u}} model. Reassuringly, this does not lead to qualitative changes to the results, but it nevertheless ensures that the grid has no impact on the generation of inner fast-rotating \emph{local} oscillation. 

We then calculate the $xy$-component of the vorticity 2-form \citep{2016PhRvD..93b4011E},
\begin{equation}\label{eq:vorticity}
  \Omega_{\mu\nu} = \nabla_{\mu}(hu_{\nu}) - \nabla_{\nu}(hu_{\mu}),
\end{equation}
in the equatorial plane. As a reminder,  $h=1+\epsilon+P/\rho_0$ represents the specific enthalpy, where $\epsilon$ is the internal specific energy and $P$ the pressure.
The time evolution plot of the $\Omega_{xy}$ contour for the Ury{\={u}} and $j$-constant models is shown in Figs. \ref{fig:uryu_vorticity} and \ref{fig:jconst_vorticity}, respectively. For the Ury{\={u}} model, starting from around 6 ms, a nonaxisymmetric oscillation develops at the inside slope of the angular velocity profile where $\Omega_{xy}$ has the lowest value (note that the dotted circle indicates the peak of the angular velocity in the Ury{\={u}} model). The instability becomes unstable and forms vortices at around 8 ms. These vortices further merge into three large vortices at around 12 ms, and eventually, merge into a large vortex in the center. For the $j$-constant model (see Fig. \ref{fig:jconst_vorticity}), a persistent spiral pattern develops around the centre throughout the simulation. These different instability behaviors distinguish the Ury{\={u}} model from the $j$-constant case. 

It has been found that differentially rotating barotropic flows with sufficiently strong shear layers are unstable to the formation of vortex chains (see  \citep{1968JFM....33..577B,1967JFM....29...39H,1984JAtS...41.1992N,1993PhFlA...5.1971S} for examples). The Ury{\={u}} model features a ring-shaped flow that rotates relatively faster than the rest of the remnant, in analogy to a jet flow within a rotating fluid (see e.g.\ \citet{1993PhFlA...5.1971S}). We suspect a similar shear instability mechanism plays a role in the observed formation of vortices. In the following, we will test an instability theorem to  get a glimpse of the underlying physics behind the \emph{inner local} instability.

\subsection{The instability criteria for rotating flow}\label{sec:stability}

\begin{figure}[!ht]
  \centering
    \subfloat[]{
    \includegraphics[clip,width=0.5\columnwidth]
                        {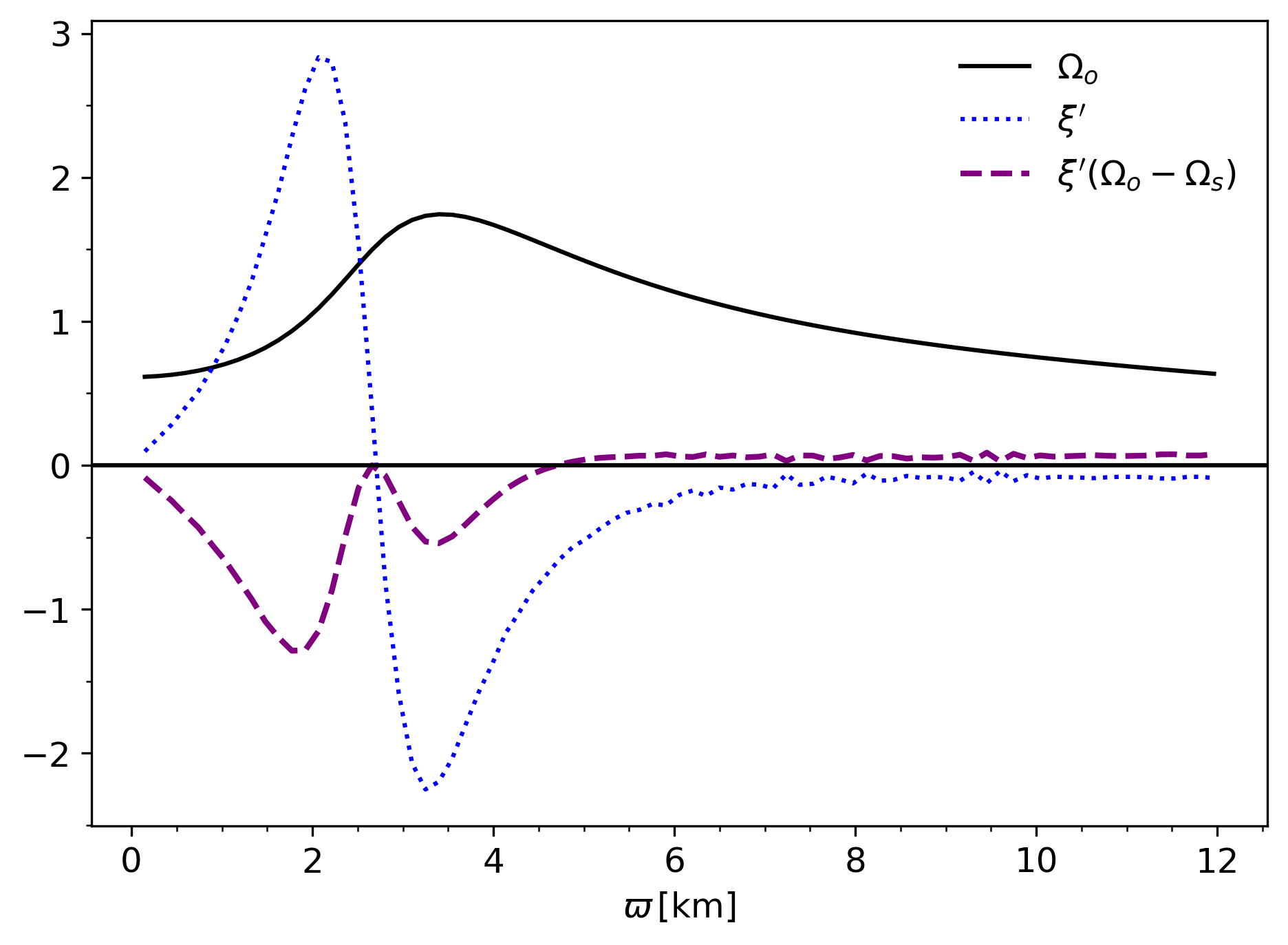}
    }
    \subfloat[]{
    \includegraphics[clip,width=0.5\columnwidth]
                    {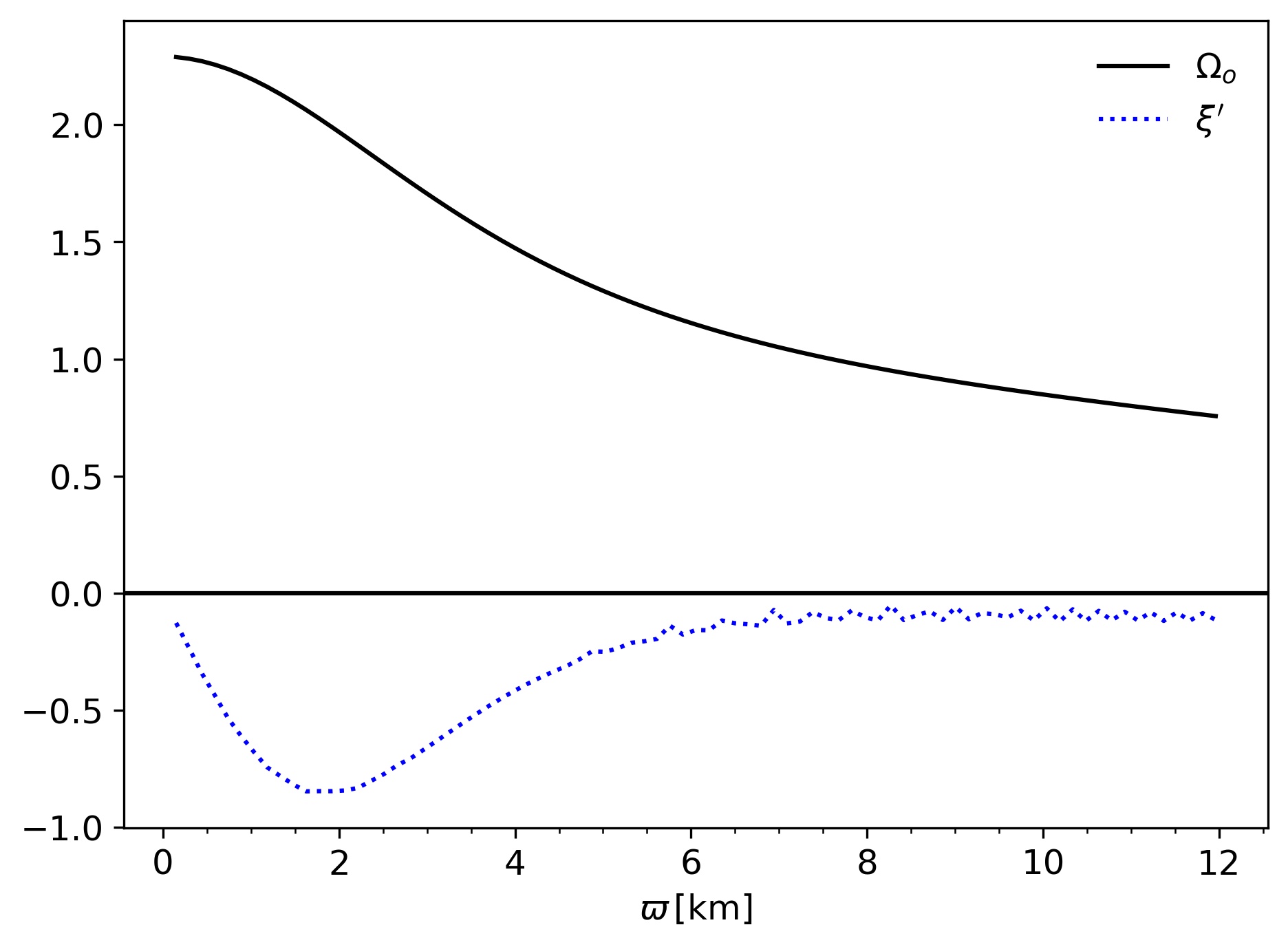}
    }
  \caption{The stability criteria plot for the Ury{\={u}} model (left panel) and the $j$-constant model (right panel). In the plot, the solid line shows the angular velocity in the equatorial plane normalized by $2\pi$, denoted as $\Omega_o$. The dotted line represents the radial gradient of the vorticity of the profile, denoted as $\xi^{\prime}$. The dashed line shows the value for the instability criterion $\xi^{\prime}(\Omega_o-\Omega_s)$ where $\Omega_s$ is the angular velocity, normalized by $2\pi$, at the inflection point $\xi^{\prime}= 0$, which only exists in the Ury{\={u}} case. Note that the region with $\xi^{\prime}(\Omega_o-\Omega_s)<0$ is expected to be unstable by the Fj{\o}rtoft criteria.}\label{fig:stability_criteria}
\end{figure}

\begin{figure}[!ht]
  \centering
    \includegraphics[clip,width=0.7\columnwidth]
                    {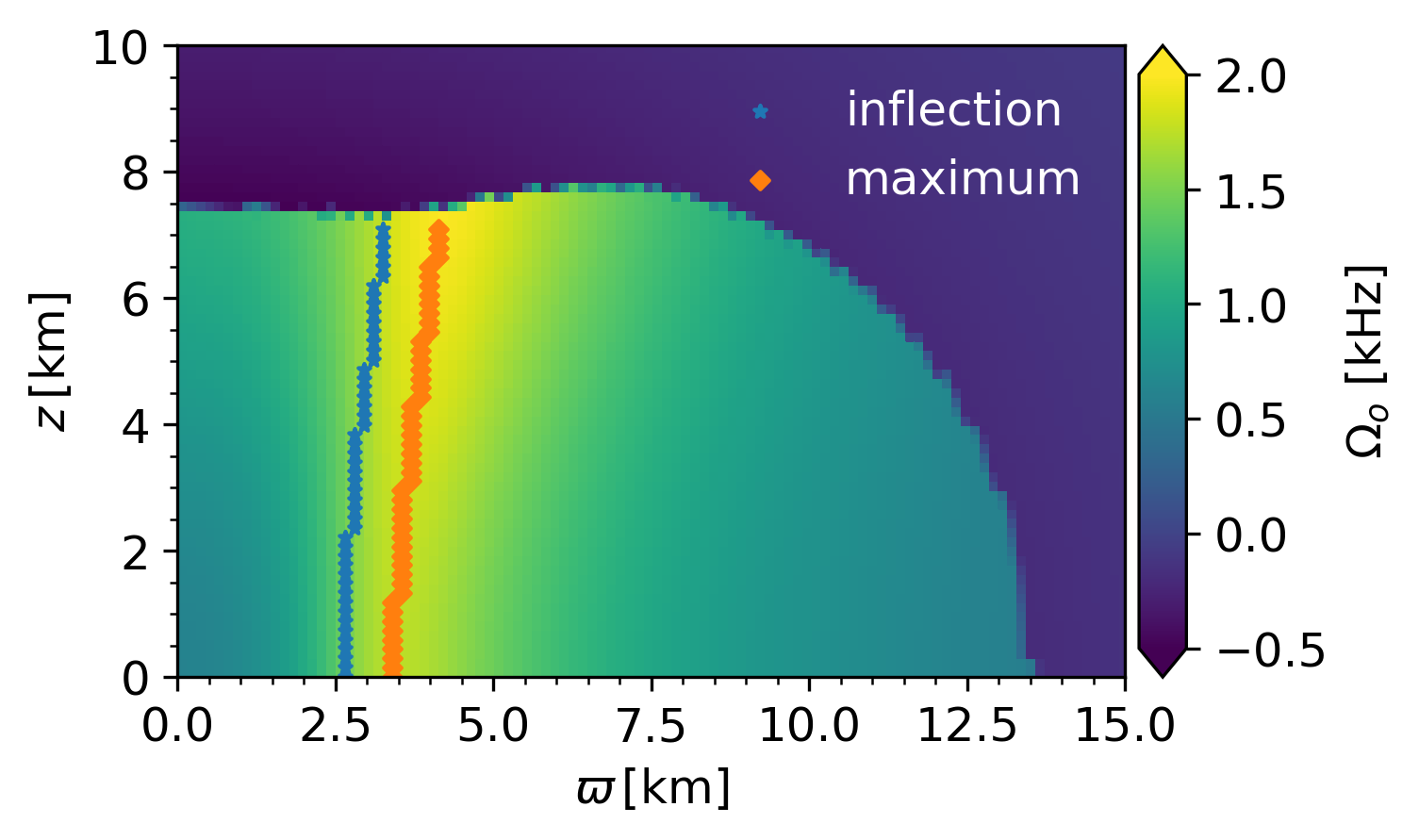}
  \caption{The normalized angular velocity $\Omega_o$ contour plot for the Ury{\={u}} model in the $\varpi$-z plane. The positions of the maximum angular velocity and its inflection point $\xi^{\prime}=0$ along the cylindrical radius $\varpi$ are also shown. \label{fig:angular_velocity_uryu}}
\end{figure}

\begin{figure*}[!ht]
  \centering
  \includegraphics[clip,width=1.0\textwidth]
                    {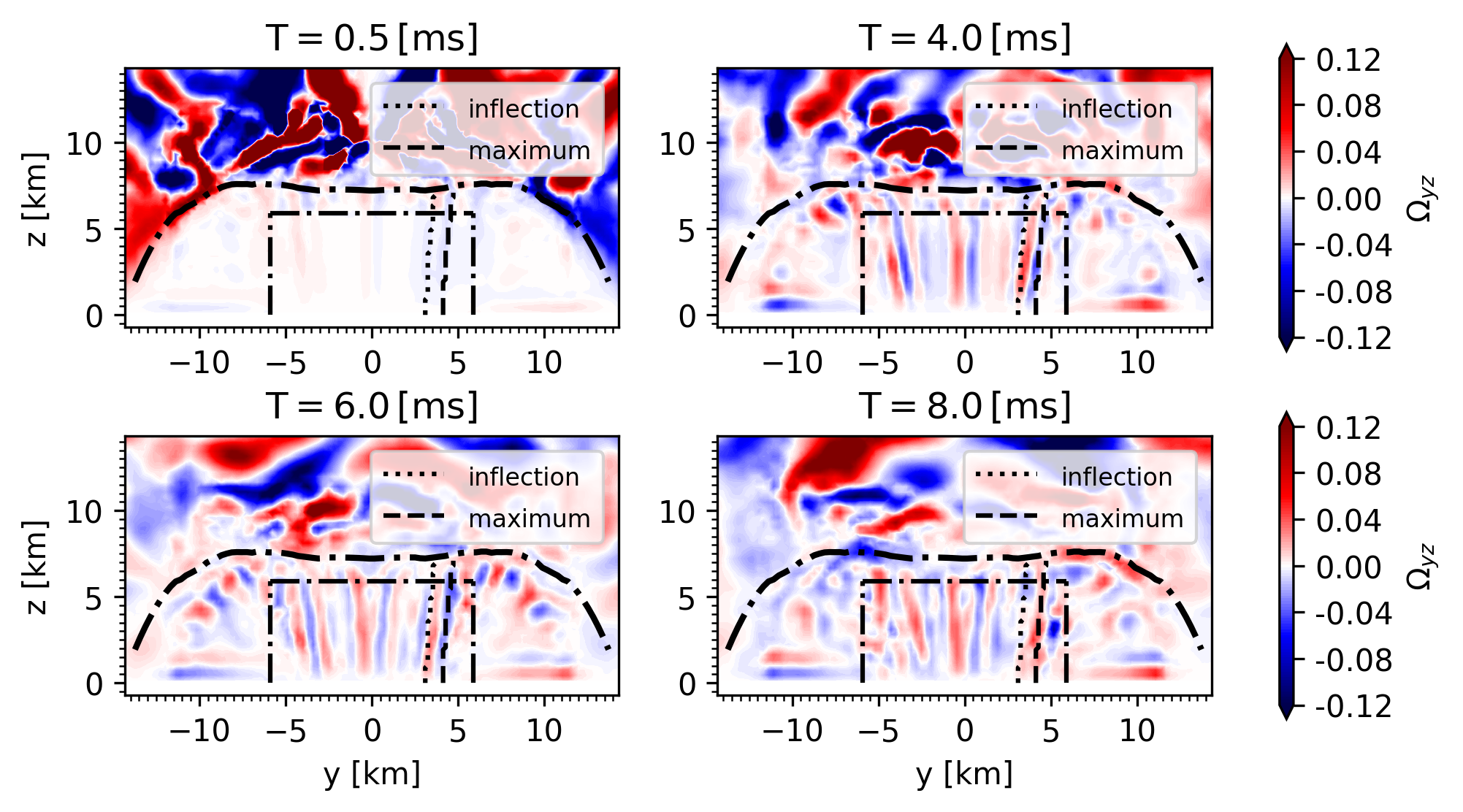}
  \caption{The time snapshots of the vorticity $\Omega_{yz}$ contour for the Ury{\={u}} model in the y-z plane. The dash and dotted line indicates the inflection point and the peak of the angular velocity, respectively. The boundary of the initial remnant profile and the refinement level is also plotted (the dotted-dash line). From the time around 0.5 ms, the shear instability around the peak region drives linear perturbation. The perturbation becomes unstable at around 8 ms. \label{fig:uryu_vorticity_yz}}
\end{figure*}

\begin{figure*}[!ht]
  \centering
  \includegraphics[clip,width=1.0\textwidth]
                    {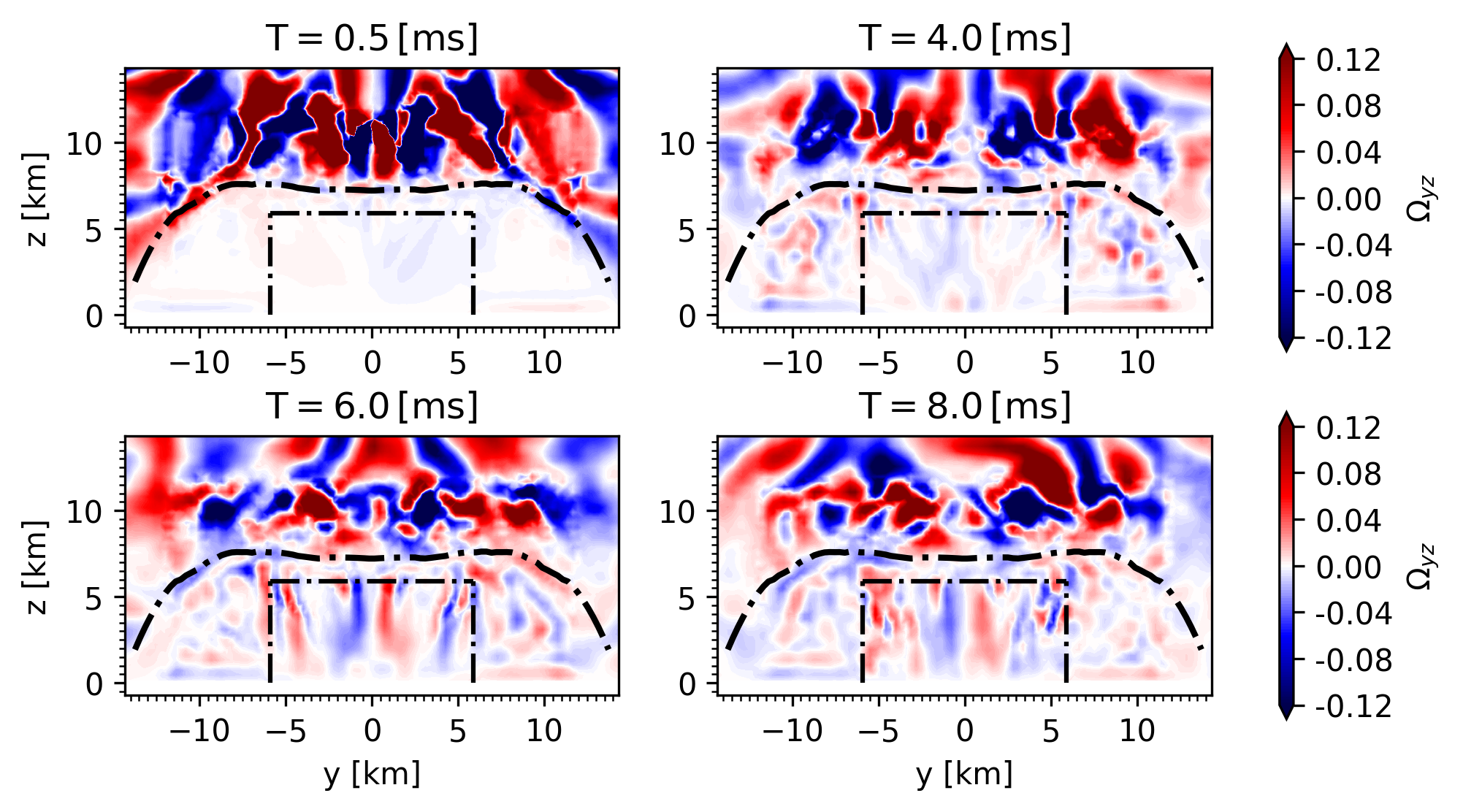}
  \caption{The time snapshots of the vorticity $\Omega_{yz}$ contour for the $j$-const model in the y-z plane. The boundary of the initial remnant profile and the refinement level is also plotted (the dotted-dash line). No meaningful physical instability is revealed in the contour plots. \label{fig:jconst_vorticity_yz}}
\end{figure*}

In the study of flow instabilities, Rayleigh first developed a general stability theory for inviscid parallel shear flows, and showed that a necessary condition for linear instability is that the velocity profile has a point of inflection \cite{doi:10.1112/plms/s1-11.1.57}. Fj{\o}rtoft then gave a strict necessary condition that there is a maximum of vorticity for inviscid instability \cite{fjortoft1950application}. For the stability of inviscid rotating flows, Rayleigh also obtained a criterion which is the analogue of the inflection point theorem in parallel flow \cite{doi:10.1112/plms/s1-27.1.5}. 

The detailed derivation of the generalization of Fj{\o}rtoft's theorem to rotating flow can be found in \cite{eckhoff1980stability}. It states that a necessary, but not sufficient, condition for linear instability of inviscid rotating flow is that $\xi^{\prime}(\Omega-\Omega_s)<0$ somewhere in the flow field. Here 
\begin{equation}
    \xi = \frac{1}{r}\frac{\partial}{\partial r}(r u_{\theta}) -\frac{1}{r}\frac{\partial u_r}{\partial \theta}
\end{equation}
is the vorticity of the background flow, $\Omega(r)$ is the mean angular velocity, and $\Omega_s=\Omega(r_s)$ is the angular velocity at the inflection point $r=r_s$ with $\xi^{\prime}=\frac{\partial \xi}{\partial r}=0$.
This argument is based on results for Newtonian barotropic fluids. In what follows we consider these criteria for a compressible relativistic fluid with minimal modification. The angular velocity used in the following analysis is defined as $\Omega_o=v/(r\,\rm{sin}\theta)/2\pi=(\Omega-\omega)e^{\beta-\nu}/2\pi$ where $v$ is the proper velocity with respect to a zero angular momentum observer and $\Omega$ is the angular velocity of the matter measured from infinity.

For general axisymmetric rotating flow in equilibrium, $u_{r}=0,u_{\theta}=\Omega(r)r$, where $\Omega(r)$ is the mean angular velocity. The vorticity of the background flow is then 
\begin{equation}
    \xi=\frac{1}{r}\frac{\partial (\Omega(r)r^2)}{\partial r}\ .
\end{equation}
We show the normalized angular velocity $\Omega_o$ and the radial gradient of vorticity $\xi^{\prime}$ for both models in the equatorial plane in Fig. \ref{fig:stability_criteria}. It is evident that the Ury{\={u}} model has one inflection point, where $\xi^{\prime}=0$, near the peak of the angular velocity profile. There is no inflection point for the $j$-constant model. Based on Rayleigh's inflection point theorem~\cite{doi:10.1112/plms/s1-11.1.57}, the rotating flow for the $j$-constant model must be stable to this linear instability, while the flow for Ury{\={u}} model can be unstable as Rayleigh's theorem represents a necessary but not sufficient condition.

From the angular velocity at the inflection point $\Omega_s$ for the Ury{\={u}} model, we further plot the instability criteria for the generalized Fj{\o}rtoft's theorem. We find there are two regions in the flow where $\xi^{\prime}(\Omega_o-\Omega_s)<0$.  The flow in these regions is more likely to be unstable based on this theorem.  Note that Fj{\o}rtoft's theorem (\cite{fjortoft1950application}) is also a necessary, but not sufficient, condition for linear instability. We suspect that a wide range of parameter values for the Ury{\={u}} model will have similar instability condition results, but currently have not determined the exact point where instability sets in.

Figure \ref{fig:stability_criteria} shows the instability criteria results in the equatorial plane. We find the instability criteria  also extend to the vertical plane. We plot the angular velocity contour in the $\varpi-z$ plane, and indicate the locations for the inflection points and the maximum angular velocity along the $\varpi$ direction (as in Fig. \ref{fig:angular_velocity_uryu}). As we can see, at each horizontal plane with different values of $z$, there exists an inflection point in the  region inside the maximum angular velocity. Instabilities similar to what we see in Fig. \ref{fig:uryu_vorticity} should thus be expected to occur at each level of $z$. In analogy to Taylor-Couette flow, the instability is essentially three dimensional. In Figs. \ref{fig:uryu_vorticity_yz} - \ref{fig:jconst_vorticity_yz}, we plot the time snapshots of vorticity $\Omega_{yz}$ in the y-z plane for the Ury{\={u}} and $j$-constant models.

In Fig. \ref{fig:uryu_vorticity_yz}, we also over-plot the vertical positions for the inflection point and maximum angular velocity of the Ury{\={u}} model. At the very beginning of the simulation, an instability develops around the  region where $\xi^{\prime}(\Omega_o-\Omega_s)<0$. The instability region almost lie along the inflection line. Until $t=6\, \rm{ms}$, the instability is in the linear regime so the shape of the instability contour does not change. At $t=8\,\rm{ms}$, the vertical instability line gets disrupted. We suspect that by this time the instability reaches the nonlinear regime (which is also suggested by Fig. \ref{fig:uryu_vorticity}). Figure~\ref{fig:jconst_vorticity_yz} shows the $\Omega_{yz}$ plot for the $j$-constant model. In comparison with the Ury{\={u}} model, we do not see clear evidence for a linear instability throughout the early phase of the time evolution in the $\Omega_{yz}$ plot. This is in agreement with our previous stability criteria results which suggest that the flow in the $j$-constant model is stable.

\subsection{Effects of grid resolution on the inner local instability}\label{sec:resolution}

\begin{figure*}[!ht]
  \centering
  \includegraphics[clip,width=1.0\textwidth]
                    {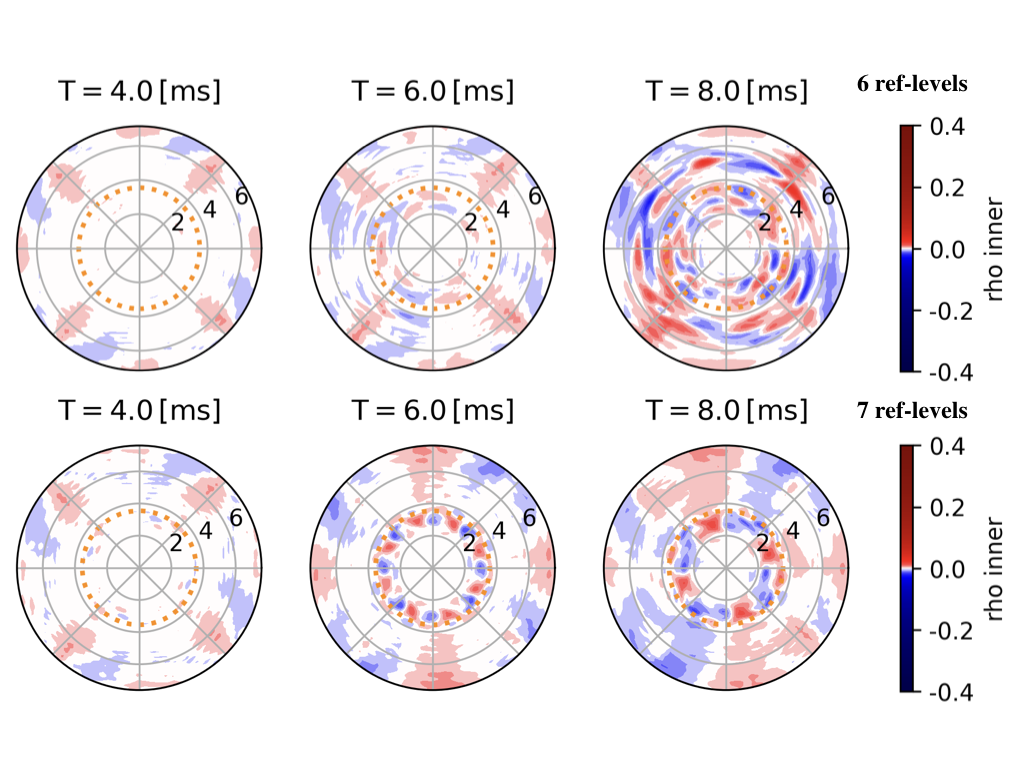}
  \caption{The early time snapshots of the density deviation contour for the Ury{\={u}} model in the x-y plane with different grid setup. The top row represents results from a simulation with six refinement levels. The circle indicates the radius of the additional refinement level which is added in the simulation with 7 refinement levels (7 ref-levels). The results corresponding to 7 ref-levels simulation are shown in the bottom row. With the addition of a finer refinement level in the center, the \emph{inner local} instability better represents itself.\label{fig:density_ref_comparison}}
\end{figure*}

In previous subsections (\ref{sec:inner} and \ref{sec:stability}), the instability for the Ury{\={u}} model develops on a uniform grid with grid spacing $dx=147\,m$. To study the effects of grid resolution, ideally we need to decrease the grid spacing for the whole simulation domain. This is computationally prohibitive. Instead, we choose to add an additional refinement level that covers the inner region where the instability occurs. With the addition of this new refinement level, the grid spacing decreases to $dx=74\,m$. In Fig. \ref{fig:density_ref_comparison}, we visualize the density deviation at early stages for both simulations. Clearly, the \emph{inner local} instability does depend on the numerical resolution. With a resolution at the sub-$100\,m$ level, the density oscillation pattern associated with the instability clearer reveals itself. From $t=6\,\rm{ms}$ to $t=8\,\rm{ms}$, the density oscillations change as discussed in Sec. \ref{sec:dynamics}. Without this refinement level, the density oscillation associated with the instability still appears but in a more subtle way.

\subsection{Low-$T/|W|$ instability and corotation points}

\begin{figure}[!ht]
  \centering
    \includegraphics[clip,width=0.7\columnwidth]
                    {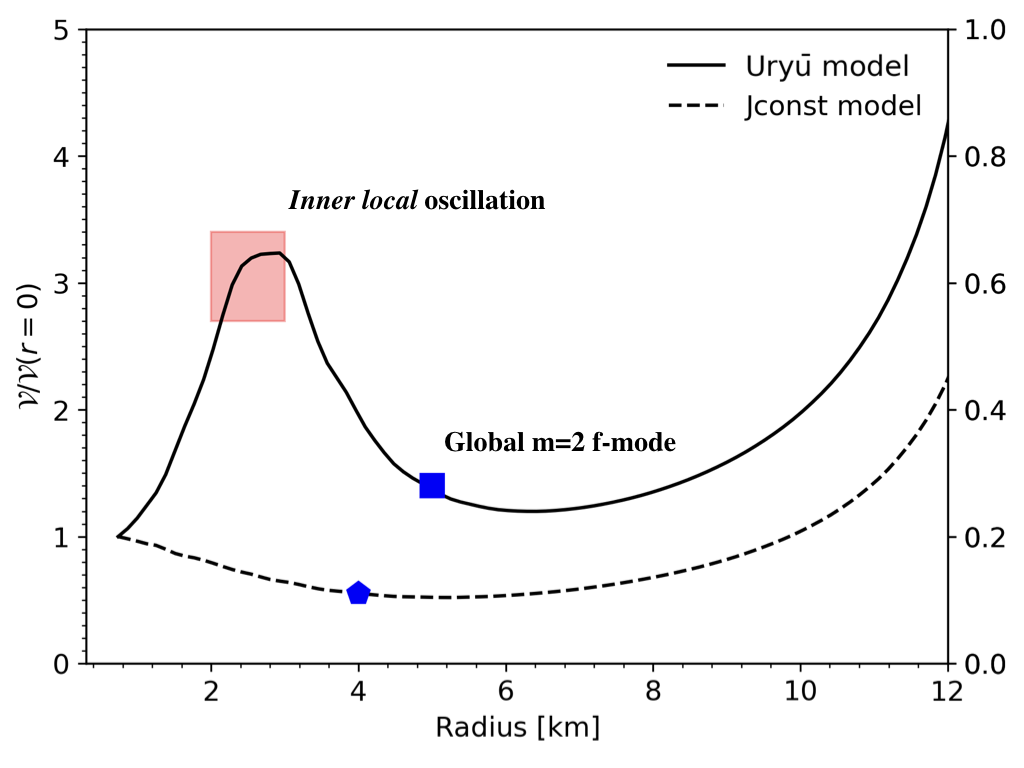}
  \caption{The radial profile of the vortensity $\mathcal{V}$ for the Ury{\={u}} model (solid line) and the $j$-constant model (dashed line). The different blue symbols represent the actual position of the corotation radii for the global $m=2$ mode frequency in each model. The red rectangle shows the corotation region for the \emph{inner local} oscillation within the Ury{\={u}} model.} \label{fig:vortensity}
\end{figure}

Now let us switch our attention to the general low-$T/|W|$ instability. The linear analysis of \citet{2005ApJ...618L..37W}  suggested that low-$T/|W|$ instabilities are triggered when the corotating $f$-mode enters the corotation band within the differentially rotating  star. By investigating the distribution of the canonical angular momentum, \citet{2006MNRAS.368.1429S} found that the  instability sets in around the corotation radius of the star, and grows as there is an inflow of angular momentum inside the corotation radius. In this picture, the corotation radius has a crucial role in that a wave propagating radially across it can be amplified. However, significant growth of a wave typically requires many passages through the corotation point \cite{2000ApJ...533.1023L}. It appears as though a resonant cavity is required to drive the modes in corotation to large amplitude \cite{2006ApJ...651.1068O}. For example, in the case of the Papaloizou-Pringle instability (PPI), the inner and outer edges of the disk or torus forms a resonant cavity in which waves are reflected back and forth \cite{1984MNRAS.208..721P}.

Similarly, \citet{1999ApJ...513..805L} have analyzed the so-called Rossby wave instability (RWI) in Keplerian accretion disks and found that it occurs when there is an extremum in the radial profile of $\mathcal{L}(r)\equiv(\Sigma \Omega/\kappa^2)S^{2/\Gamma}$, where $\Sigma$ is the surface mass density of the disk, $\Omega$ is the angular rotation rate, $S(r)$ is the specific entropy, $\Gamma$ is the adiabatic index, and $\kappa$ is the radial epicyclic frequency. Extrema of $\mathcal{L}(r)$ could come from several sources. In \cite{1999ApJ...513..805L} they considered the special case where there is a local maximum in the disk entropy profile, $S(r)$. This maximum acts to trap the waves in the vicinity of the maximum, given sufficiently strong variation. \citet{2000ApJ...533.1023L} have presented a detailed linear theory for the RWI and show that it exists for a wider range of conditions, specifically, for the case where there is a ``jump'' over some range of $r$ in $\Sigma(r)$ or in the pressure $P(r)$. They also point out that the profiles of $\Sigma(r)$ and $P(r)$ considered are not the only ones which may lead to instability. For example, a profile with local extreme in the vortensity distribution [see Eq. \eqref{eq:vortensity}] may also give instability. Recently, \citet{2010A&A...516A..31M,2012A&A...542A...9M,2012MNRAS.422.2399M} carried out full 3D numerical simulations of the RWI in protoplanetary discs. The simulations show that the RWI can develop in 3D discs as in 2D when an extremum exists in the background fluid vortensity. 

Closely related to this, in the study of differentially rotating neutron stars, \citet{2006ApJ...651.1068O} show that models of differentially rotating neutron stars can also exhibit a local minimum in their radial vortensity profile, and a similar resonant cavity mechanism seems to trigger the one-armed spiral instability. In addition to the one-armed ($m=1$) spiral mode, they have found that higher order ($m=2$ and $3$) nonaxisymmetric modes can also become unstable if the associated corotation points that resonate with the eigenfrequencies of these higher modes also appear inside the star. Note that their model configurations feature centrally condensed, rather than toroidal, density structures. These studies suggested that the presence of a minimum in the profile of the vortensity [see Eq. \eqref{eq:vortensity}] is a necessary condition for a mode in corotation to be unstable \cite{2006ApJ...651.1068O,2010CQGra..27k4104C}.

Given the different shape of the angular velocity for the Ury{\={u}} model and $j$-constant model, we expect the vortensity profile for the two models to be different. To verify this and facilitate further analysis, we have computed for both models the Newtonian vortensity, defined as the ratio, along the radial cylindrical coordinate, between the radial vorticity and the density \cite{2000ApJ...533.1023L,2010CQGra..27k4104C}, i.e.
\begin{equation}\label{eq:vortensity}
  \mathcal{V}= \frac{\kappa^2}{2\Sigma \Omega} =\frac{2\Omega+\varpi \Omega_{, \varpi}}{\rho}.
\end{equation}
where $\kappa^2=\frac{2\Omega}{r}\frac{d}{dr}(r^2\Omega)$ is the square of the radial epicyclic frequency (so that $\kappa^2/2\Omega$ is the vorticity) \citep{2012MNRAS.422.2399M}. The results are shown in Fig. \ref{fig:vortensity}. The corotation points of the global $m=2$ mode (see Fig. \ref{fig:angular_velocity_evolution}) are also indicated. Both corotation points are located near the minimum of each vortensity profile. This is in agreement with previous studies (see e.g.\ \citep{2010CQGra..27k4104C}).

Now let us focus on the \emph{inner local} oscillation for the Ury{\={u}} model. Based on Fig. \ref{fig:eta_plus_spec}, the oscillation has an frequency of 3.6 kHz and the dominant mode azimuthal number is $m=2$ (inferred from Fig. \ref{fig:uryu_density_reflux}) during the time period $t > 10\rm{ms}$. With a pattern speed of 1.8 kHz, its corotation point is close to the peak of the angular velocity profile (see Fig. \ref{fig:angular_velocity_evolution}). During the period where the oscillation develops (roughly $t=8\,\rm{ms}{-}16\,\rm{ms}$), the angular velocity profile of the remnant changes. Still, we can roughly find the ``corotation region'' of the local mode which is close to the peak of the vortensity profile (see Fig. \ref{fig:vortensity}).   

Corotation of modes with rotational profiles of postmerger remnants has also been found in long-term merger simulations \citep{2020PhRvD.101f4052D}. In particular, \citeauthor{2018PhRvL.120v1101D} find a convective instability in the postmerger remnant. The initial convectively unstable region is located near the peak of the angular velocity profile, which is similar to what we find for the \emph{inner local instability} using the vorticity 2-form. The highest resolution achieved in \citep{2018PhRvL.120v1101D,2020PhRvD.101f4052D} is $dx=185\,\rm{m}$ for long-term simulations. In contrast, the simulations conducted in our study focus on the late postmerger phase. As we show in Sec. \ref{sec:resolution}, a resolution at the $\rm{sub}-100\,m$ level is required to resolve the instability. In this study, we discuss the analysis tools and numerical setup required to diagnose the local instability associated with postmerger remnants. We expect the tools and analysis can be brought to bear on  long-term merger simulations, as well.

\subsection{Linear and nonlinear instability for the \emph{inner local} instability}

Figure \ref{fig:uryu_vorticity} illustrates the process of the formation and development of the inner local instability within the star's core. Roughly speaking, it goes through three stages. First, the rapid growth of the initial small amplitude perturbations. Next follows the production of large-scale vortices and their interactions with the background flow. Finally there is a coupling of vortices with global spiral waves. For the first stage, roughly $t<8\,\rm{ms}$, the shear instability described in subsection \ref{sec:stability} plays a major role. Later on, a nonlinear disturbance leads to the formation of vortices and the oscillation pattern. These stages are similar to the case of the RWI in thin accretion disks with density or pressure structures (see e.g.\ \cite{2001ApJ...551..874L}). When we discuss the \emph{inner local} oscillation for the Ury{\={u}} model, the shear instability based on Rayleigh's and Fj{\o}rtoft theorem, and corotation resonance effects from RWI/PPI may be well entangled. At the end of the simulation, the \emph{inner local} oscillation almost synchronises with the outer $f$-mode oscillation.

\section{Conclusions}

We have carried out numerical simulations of rapidly and differentially rotating neutron star configurations, inspired by postmerger remnants. 
The results demonstrate that different angular velocity profiles lead to slightly different dynamics, impacting on the growth and saturation of the low-$T/|W|$ instability. In particular, we find that the profile of the Ury{\={u}} model generates a more dominant $m=2$ mode perturbation. For the $j$-constant model with similar bulk properties, the $m=2$ mode is more strongly coupled to other multipole modes (especially an $m=3$ component). In this case, the mode coupling generates distinct gravitational-wave bursts rather than  a continuously growing amplitude in the beginning.

In an addition, we find the oscillation pattern close to  the center of the remnant behaves differently in the Ury{\={u}} model, a feature that triggers a \emph{local} instability. We show that this \emph{local} instability is directly linked to the bell-shaped feature of the angular velocity profile, and occurs in the inner part of the remnant where strong shear layer exists. We apply the generalized  Fj{\o}rtoft's theorem to the rotating profiles, and find that the Ury{\={u}} model has inflection points in its angular velocity profile which satisfy the instability criteria, while the $j$-constant model appears to be stable according to this measure. The vorticity contour of the Ury{\={u}} model confirms that the \emph{inner local} instability  occurs in the predicted unstable region.  This \emph{inner local} instability starts with a linear instability in the horizontal as well as in the vertical plane. It then leads to the formation of vortices, which  merge together to form a fast rotating m=2 oscillation, distinguishable from the global m=2 mode that appears in the rest of the remnant. As time goes on, the \emph{inner local} $m=2$ oscillation synchronizes with the global $m=2$ mode. For the $j$-constant stellar model, we only observe the global mode development. 

For the general low-$T/|W|$ instability, we find that an $m=2$ f-mode corotation point exists inside the rotating profile in both cases. This  corotation point is located near the minimum of the corresponding vortensity profile. This indicates that a corotation resonance may amplify the magnitude of the f-mode as discussed in the case of RWI/PPI. For the Ury{\={u}} model, the corotation point for the \emph{local} $m=2$ oscillation is located near the peak of the angular velocity profile. The linear shear instability and RWI/PPI may well participate in the development of this \emph{inner local} oscillation.

This study provides an initial survey of the nonlinear effects associated with unstable modes for different rotation laws, complementing the linear perturbation study from \cite{2020arXiv200310198P}. We focused on comparing a rotation profile inspired in binary neutron star merger remnants to the standard $j$-constant rotation law. The results provide qualitative insights into the impact of the rotation profile on the development of mode instabilities. 
We have performed some additional simulations with the same models, but with additional parameters chosen so that no corotation points exist, and these show no instability. However, we have not found the instability threshold to high accuracy.
A more detailed parameter survey, exploring the  dependence on the different parameters, like the  peak and position of the angular velocity, may lead to a deeper understanding about the instability and mode dynamics during the postmerger phase.
The discovery of the \emph{inner local} instability highlights the importance of the study of instabilities of shear flows in the  framework of relativity. In this study, we adopted a clean setup, minimizing the number of variables for the comparison. For a more realistic setup, additional features  need to be considered. First of all, HMNSs found in merger simulations have complex microphysical equations of state and temperature profiles. These two factors impact on the sound speed, and the pressure support of the HMNSs, and  may thus influence the transfer of energy and angular momentum within the rotating profiles. Also, we ignored (at least initially) the extended disk surrounding the HMNS. The matter is this disk probably accretes onto the rotating core of the HMNS  on a dissipative time scale which may affect the long-term stability of the remnant. These effects all require further investigation and we may return to the problem in the future.

The initial data, parameter file, analysis and visualization scripts are available through Zenodo \citep{xiaoyi_xie_2020_3860828}. 

\acknowledgments
The authors acknowledge the use of the IRIDIS High Performance Computing Facility, and associated support services at the University of Southampton, in the completion of this work. IH and NA gratefully acknowledges financial support from STFC via Grant No. ST/R00045X/1.

\bibliography{rnsid.bib}
\bibliographystyle{apsrev4-1}

\end{document}